\documentclass[fleqn,10pt]{wlscirep}
\usepackage[utf8]{inputenc}
\usepackage[T1]{fontenc}
\usepackage{amsmath}
\usepackage{dsfont}
\usepackage{layouts}
\usepackage{subcaption}
\title{Training the parametric interactions in an analog bosonic quantum neural network with Fock basis measurement}

\author[1]{Julien Dudas}
\author[1]{Baptiste Carles}
\author[2]{Elie Gouzien}
\author[1]{Julie Grollier}
\author[1, *]{Danijela Marković}
\affil[1]{Laboratoire Albert Fert, CNRS, Thales, Université Paris-Saclay, 91767 Palaiseau, France}
\affil[2]{Alice \& Bob, 53 Bd du Général Martial Valin, 75015 Paris, France}

\affil[*]{danijela.markovic@cnrs-thales.fr}

\begin{abstract}

Quantum neural networks promise to extend the power of machine learning into the quantum domain, with potential applications ranging from automatic recognition of quantum states to the control of quantum devices. However, their physical implementation and training remain challenging. In particular, the backpropagation algorithm that underpins the efficiency of classical neural networks cannot generally be applied to large quantum systems, as nonlinear quantum dynamics are not efficiently simulable. Instead, variational quantum circuits typically rely on parameter-shift rules or sampling-based gradient estimation.
Here we propose a bosonic quantum neural network based on parametrically coupled Gaussian modes. Although the underlying quantum dynamics are linear, nonlinear output features are generated through Fock-basis measurements. Because Gaussian evolution can be efficiently simulated in the Heisenberg representation, the system admits gradient-based optimization by differentiating a classical model of the dynamics, while the forward evolution itself could be implemented on quantum hardware. This hybrid approach enables end-to-end training of physically meaningful parameters without requiring gradient extraction from the experimental device.
Such architectures are naturally compatible with circuit quantum electrodynamics platforms featuring tunable parametric couplers, as well as integrated photonic systems with engineered $\chi$(2) or $\chi$(3) nonlinearities. Our results demonstrate that linear bosonic networks combined with nonlinear measurement provide a scalable and trainable route toward experimentally realizable quantum neural networks.

\end{abstract}

\begin{document}
\flushbottom
\maketitle

\section*{Introduction}

The potential of quantum systems for computation has long been recognised, rooted in their ability to exist in superposition and entangled states. These properties suggest that quantum computers may outperform classical systems for certain tasks that exploit quantum parallelism~\cite{groverFastQuantumMechanical1996, shorPolynomialTimeAlgorithmsPrime1997, harrowQuantumAlgorithmLinear2009}. In a similar spirit, quantum neural networks (QNNs) aim to leverage quantum dynamics to generate feature representations that may be difficult to reproduce efficiently with classical models. They are particularly attractive when learning directly from quantum data or when interfacing with other quantum devices~\cite{huang_quantum_2022}.

A central challenge, however, lies in their training. The backpropagation algorithm that underpins the success of classical neural networks and large language models relies on the ability to efficiently simulate system dynamics and compute gradients. While this approach can be extended to some classical physical neural networks — where the forward pass is performed on hardware and gradients are computed using a tractable model — it is generally not applicable to large quantum systems, whose generic many-body dynamics cannot be efficiently simulated classically. As a result, most quantum neural network proposals either avoid training the physical system altogether, as in quantum reservoir computing~\cite{fujii_harnessing_2017, senanianMicrowaveSignalProcessing2024, carles_experimental_2025, dudas_quantum_2023, govia_quantum_2021, nokkalaGaussianStatesContinuousvariable2021}, or rely on alternative gradient-estimation techniques such as parameter-shift rules in variational quantum circuits~\cite{schuld_evaluating_2019, mitarai_quantum_2018} . Reservoir computing minimizes training overhead but keeps the internal dynamics fixed, limiting its ability to tailor feature representations to a specific task, while parameter-shift approaches can become resource-intensive as system size grows~\cite{benedetti_parameterized_2019, larocca_barren_2025}.

In this work, we propose a bosonic quantum neural network based on parametrically coupled Gaussian modes, with nonlinear output features obtained through Fock-basis measurements. This approach combines two key ingredients: (1)~Gaussian dynamics are linear and can be efficiently simulated in the Heisenberg representation, enabling gradient computation by automatic differentiation of a classical model of the system; (2)~nonlinear features arise at the measurement stage through Fock-state statistics. Although our study is fully numerical, the forward dynamics could be implemented using platforms with tunable parametric couplings, such as circuit quantum electrodynamics architectures~\cite{metelmannParametricCouplingsEngineered2023} or integrated photonic systems with engineered $\chi$(2) or $\chi$(3) interactions~\cite{boydNonlinearOptics2008}, while the backward pass is performed on a classical model. This hybrid strategy, where the physical system performs the forward computation and gradients are computed on a differentiable model, corresponds to a form of physics-aware training that has previously been implemented in classical neuromorphic networks~\cite{wright_deep_2022}.

\begin{figure*}[htp]
    \centering
    \includegraphics[width=\linewidth]{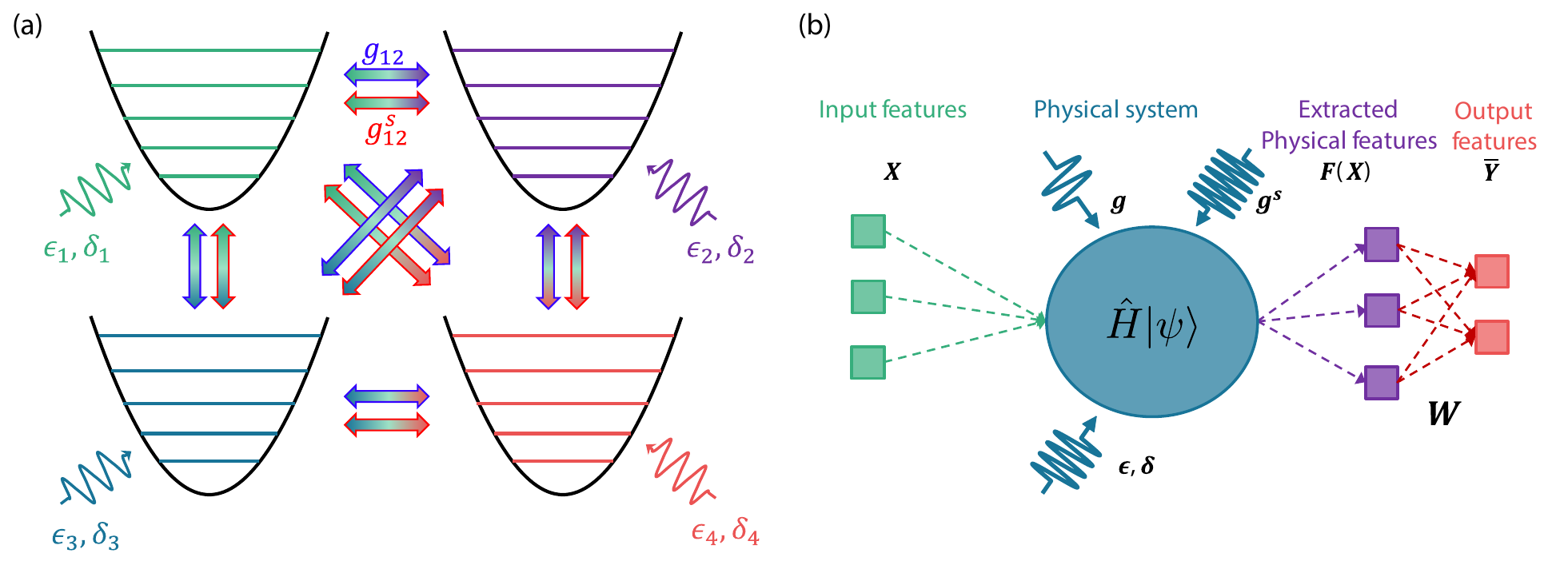}
    \caption{\textbf{Analog quantum bosonic neural network.} (a) A set of bosonic modes (here 4), driven close to resonance with detunings $\delta_k$ and amplitudes $\epsilon_k$, and with dissipation rates $\kappa_k$. Two types of coupling processes are driven, photon exchange at a rate $g_{kl}$ (blue-outlined arrow) and two-mode squeezing at a rate $g^s_{kl}$ (red-outlined arrow). (b) Schematic of an analog quantum neural network. Input data vector $\bf{X}$ (green squares) is encoded into drive parameters, and feature vector $\bf{F(X)}$ (purple squares) is obtained by measuring probabilities $P_k(n)$ of a mode $k$ to contain $n$ photons. Prediction $\bf{\bar{Y}}$ (red squares) is obtained by multiplying the feature vector by a trained weight matrix $\bf{W}$. }
    \label{fig:Figure1}
\end{figure*}

In the present work, we focus on parametric processes arising from three-wave mixing and described by quadratic Hamiltonians, whose Gaussian dynamics can be efficiently simulated in the Heisenberg representation. Although the evaluation of Fock-state measurement probabilities introduces additional computational overhead, the framework remains tractable in the parameter regimes explored here. We demonstrate learning on three representative tasks: time-series processing, highly nonlinear data classification, and higher-dimensional image classification. For each task, the number of measured Fock-state probabilities is treated as a computational resource and chosen to be the minimal number required to achieve high performance, increasing only with task complexity. This setting enables a direct comparison between trained and untrained Gaussian architectures, allowing us to assess the impact of parameter optimization on computational performance.

\section*{Results}

We consider a set of $M$ modes pairwise coupled through two parametric processes: coherent photon conversion at a rate $g_{kl}$ and two-mode squeezing at a rate ${g^s}_{kl}$ for modes $k$ and $l$ (Figure~\ref{fig:Figure1}a). In the rotating frame, the Hamiltonian of this system writes
\begin{equation}
\label{eq:Hamiltonian} 
\left\{ 
\begin{aligned} 
\hat{H} &= \hat{H}_0 + \hat{H}_{\rm{in}} \\ \frac{\hat{H}_0}{\hbar} &= - \sum_{k=1}^{M} \delta_k \hat{a}^\dagger_k \hat{a}_k + \sum_{k<l} g_{kl} \hat{a}_k^\dagger \hat{a}_l + {g^s}_{kl} \hat{a}_k^\dagger \hat{a}_l^\dagger + \mathrm{h.c.} \\ \hat{H}_{\rm{in}} &= i \hbar \sum_{k} \sqrt{\kappa_k}\hat{a}_k \hat{a}_{k, \rm{in}}^\dagger + \mathrm{h.c.} 
\end{aligned} 
\right. 
\end{equation}
where $\hat{H}_0$ and $\hat{H}_{\text{in}}$ are respectively the Hamiltonian of coupled modes and the drive Hamiltonian, $\delta_k$ is the drive detuning of the mode $k$ from its resonance frequency and $\kappa_k$ is its coupling rate to the transmission line. We neglect here the internal losses of the modes. The input modes $\hat{a}_{k, \text{in}}$ are coherent states of complex amplitude $\epsilon_k = \langle \hat{a}_{k, \text{in}} \rangle $.

We train two layers of weights, as shown in Figure~\ref{fig:Figure1}b. The first layer is composed of the complex drive parameters, that is, the amplitudes, phases, and detunings of the nearly resonant drives, as well as the amplitudes and phases of the coupling tones. The second layer is composed of the output weights $\bf{W}$. Both of these layers of weights are jointly trained through backpropagation. Detunings of the coupling tones are not free parameters, as in the rotating wave approximation, they are only efficient if they are set to ${\delta^s}_{kl} = \frac{1}{2}(\delta_k + \delta_l)$ for the two-mode squeezing tone and $\delta_{kl} = \frac{1}{2}(\delta_k - \delta_l)$ for the coherent photon conversion tone. All physical parameters can be represented as vectors: $\bf{\epsilon}$ stores the nearly resonant drive amplitudes, $\bf{\delta}$ the detunings, $\bf{g}$ the photon conversion rates, $\bf{g^s}$ the two-mode squeezing rates and $\bf{\kappa}$ the transmission line coupling rates. Depending on the task, we choose to encode the input data $\bf{x}$ in the amplitude or phase of one of these vectors of parameters, that we now call $\bf{\theta}$, using the encoding
\begin{equation}
        \label{eq:encoding}
        \bf{\theta}(\bf{x}) = \bf{\theta}^T_0  \bf{x} + \bf{\theta}_{\text{bias}}.
    \end{equation}
We train all the other complex parameters $\boldsymbol{\theta}$, output weights $\mathbf{W}$, as well as the encoding range determined by the prefactor $\boldsymbol{\theta_0}$ and bias $\boldsymbol{\theta_{\text{bias}}}$. Because the exchange coupling rates are constrained to be real, which will later be justified, the number of trainable physical parameters is $\frac{3}{2}M(M-1)+3M$ (see Section VI.A in the Supplementary Material) and scales quadratically with the number of coupled modes $M$, while the dimension of the underlying Hilbert space scales exponentially. For all tasks considered in this work, the input data are rescaled to lie within the interval $[0,1]$. This normalization facilitates controlled tuning of the encoding parameters, ensuring that their absolute values remain bounded by the encoding range defined by $\boldsymbol{\theta_0}$ and $\boldsymbol{\theta_{\text{bias}}}$.

Keeping experimental feasibility in mind, we choose to measure local Fock-state probabilities rather than joint ones. Importantly, we find that this does not reduce the performance of the quantum neural network, while significantly reducing the computational cost associated with measurement evaluation. For Gaussian states, photon-number probabilities can be obtained by evaluating the overlap between the Gaussian density operator and Fock states. Using the phase-space representation of Gaussian states, such as Husimi Q-function formalism, this overlap can be expressed analytically in terms of the covariance matrix and displacement vector, leading to loop hafnian expressions for displaced Gaussian states~\cite{hamiltonGaussianBosonSampling2017}. We use this result to evaluate local Fock-state probabilities
\begin{equation} \label{eq:GBS} P_k(n|\bf{\alpha}, \bf{\sigma}) = \frac{\exp(-\frac{1}{2} \bf{\alpha}_k^\dagger \bf{\sigma}_{k,Q}^{-1} \bf{\alpha}_k )}{n! \sqrt{\det(\bf{\sigma}_{k,Q})}} \text{lhaf}(\bf{A}_n), \end{equation} where \begin{equation} \label{eq:GBS_intermediate_matrices_local} \begin{cases} \bf{\sigma}_{k,Q} & = \bf{\sigma}_k + \frac{1}{2}\mathds{1}_2\\ \bf{T} & = \begin{pmatrix} 0 & 1 \\ 1 & 0 \end{pmatrix}\\ \bf{A} & = \bf{T} \left( \mathds{1}_2 - \bf{\sigma}_{k,Q}^{-1} \right) \end{cases}. \end{equation}
$\boldsymbol{\alpha}_k$ and $\boldsymbol{\sigma}_k$ are the displacement vector and covariance matrix of mode $k$. The matrix $\boldsymbol{A}_n$ is constructed from $\boldsymbol{A}$ by substituting its diagonal with $\bf{\alpha}_k^\dagger \bf{\sigma}_{k,Q}^{-1}$ and repeating rows and columns according to the detected photon number $n$. The loop hafnian $\mathrm{lhaf}(\cdot)$~\cite{bjorklundFasterHafnianFormula2019} introduces a computational cost that scales exponentially with the number of detected photons. In this work, photon numbers are explicitly constrained during training and measurements are restricted to low photon numbers ($n<10$), ensuring that the evaluation of local Fock-state probabilities remains tractable in practice for the system sizes considered here. 

In our simulations, gradients are obtained by automatic differentiation of this classical probability expression within PyTorch, and parameters are optimized using the Adam algorithm~\cite{kingmaAdamMethodStochastic2017}, based on code adapted from Ref.~\cite{bulmerBoundaryQuantumAdvantage2022}. In a potential experimental implementation, output probabilities would be measured on a Gaussian boson sampling device during the forward pass, while gradients would be computed from the same differentiable classical Gaussian model used here.

\subsection*{Sine and square waveform classification}

\begin{figure}[htp]
    \centering
    \includegraphics[width=0.5\linewidth]{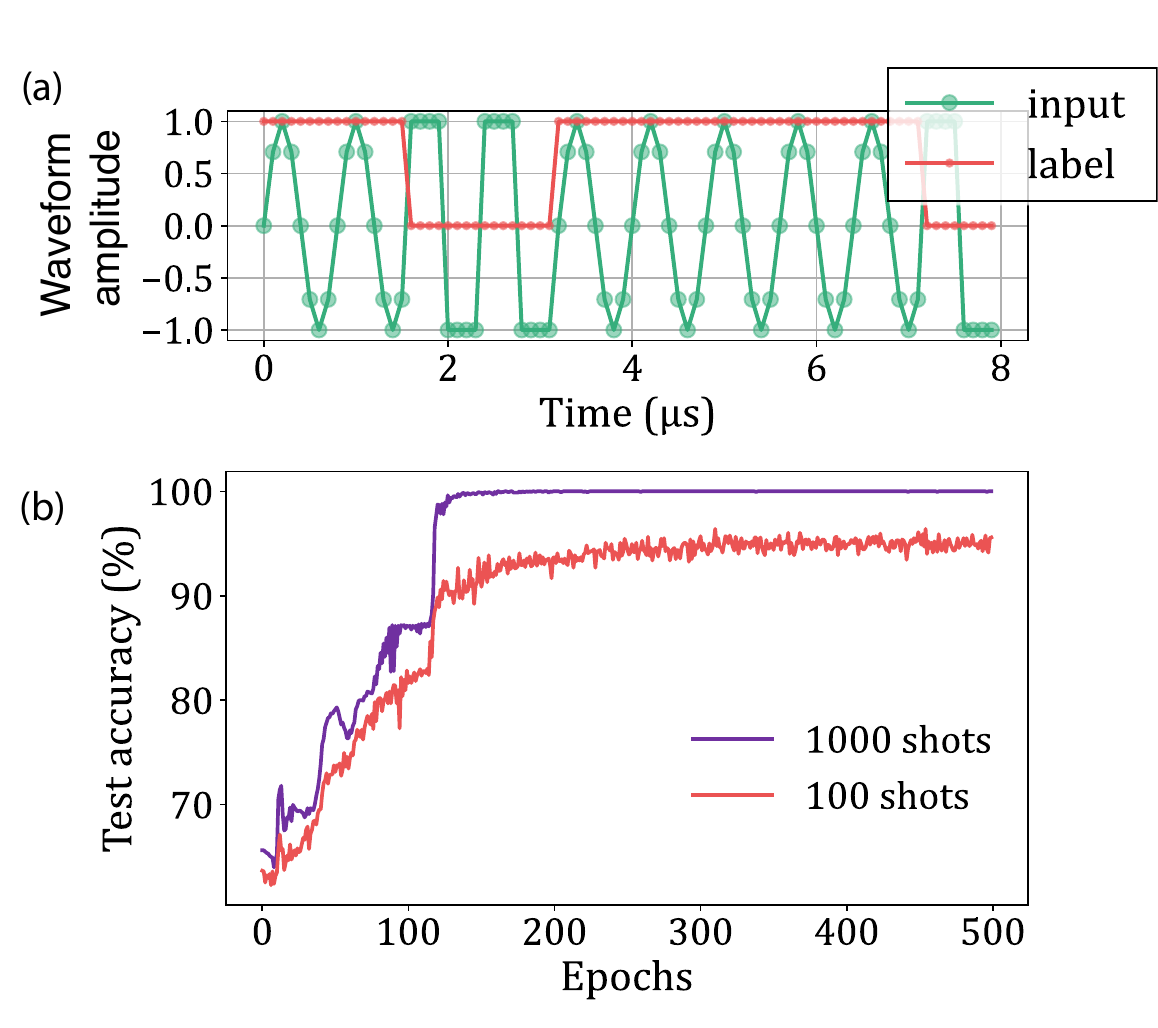}
    \caption{\label{fig:fig_sin} \textbf{Sine/square classification task.} (a) The input data is a time series of waveform amplitudes corresponding to randomly generated sine or square signals, each discretized into 8 time steps. The vertical axis represents the value of the waveform at each discrete time step. The task consists in predicting to which waveform each point belongs. The label, shown in red, is 1 for points belonging to a sine waveform and 0 for points belonging to a square waveform. (b)~Accuracy as a function of the number of training epochs $N_{\rm{epochs}}$ for two different numbers of measurement shots $N_{\rm{shots}} \in \{100, 1000\}$ used to determine the probability $P_1(0)$.}
\end{figure}

We first demonstrate learning of the sine and square waveform classification. This dataset, shown in Figure~\ref{fig:fig_sin}(a), consists of 200 randomly generated sine and square waveforms, each discretized into 8 sample points and presented sequentially to the network. Each individual sample point is treated as a datapoint and labeled according to whether it originates from a sine (class 1) or square (class 0) waveform. This yields a total of 1600 labeled datapoints, which are split into 800 training and 800 test samples.
For this task, we use two coupled modes. Classical input values $\bf{x}$ are mapped to the complex amplitudes of their nearly resonant drives $\mathbf{\epsilon}$ using the encoding map Eq.~\ref{eq:encoding}. The drive detunings, coherent coupling and two-mode squeezing rates are learned, as well as the prefactor and bias for the drive amplitudes. Each input point is sent for a duration $\delta t = \frac{2 \pi}{5 \kappa}$, where $\kappa$ is the average of the dissipation rates. The drive amplitudes are limited to a range that ensures negligible probability amplitudes for photon states higher than 9. The loss function applied for this task is the mean square error (MSE), 
\begin{equation}
    f(\mathbf{\bar{Y}}, \mathbf{Y}) = \frac{1}{N} \sum_{i=1}^N (\mathbf{\bar{Y_i}} - \mathbf{Y_i})^2,
\label{MSE}
\end{equation}
where $N$ is the number of data points used for training, $\bf{\bar{Y}} = \bf{W F(X)} $ is the network prediction obtained by multiplying the feature vector $\bf{F(X)}$ containing the measured probabilities by the weights matrix $\bf{W}$, and $\bf{Y}$ is the target vector. Parameter values are constrained to ranges typically used in quantum superconducting circuits~\cite{blaisCircuitQuantumElectrodynamics2021}.
To prevent the training from pushing the parameters to values that cause photon numbers to diverge, we introduce a regularization to the loss function, 
\begin{equation}
    \text{loss}_{\langle \bar{N} \rangle} = 
        \beta_{\langle \bar{N} \rangle} \times \text{MSE}\left(\langle \bar{N} \rangle_{\text{avg}}, \langle \bar{N} \rangle_{\text{tg}}\right),
\label{regularization}
\end{equation}
where $\langle \bar{N} \rangle_{\text{avg}}$ is the average of the photon number expectation values $\langle \bar{N} \rangle$ over the time interval $\delta t$, for the maximal and the minimal valued input. The target average photon number $\langle \bar{N} \rangle_{\text{tg}}$ is set to 2 photons in each mode to ensure that the occupations of measured Fock states are non-negligible while at the same time not over constraining the learning range. The parameter
$\beta_{\langle \bar{N} \rangle}$ is a prefactor that controls the influence of $\text{loss}_{\langle \bar{N} \rangle}$ on the overall learning process. The total optimized loss function is then $f(\bf{\bar{Y}}, \bf{Y}) + \text{loss}_{\langle \bar{N} \rangle}$. This regularization does not model photon loss or dissipation, rather, it biases the optimization toward low-excitation regimes where the Gaussian description remains accurate, numerical truncation effects are negligible, and experimental implementations are more stable.

The results are summarized in Table~\ref{tab:sinsquare}. To isolate the effect of training the physical parameters, we compare the performance of the bosonic QNN to the same hardware operated as an untrained feature generator, i.e., in a quantum reservoir computing configuration~\cite{dudas_quantum_2023}. In both cases, the classical input is encoded in the amplitudes of the nearly resonant drives. In the reservoir configuration, however, all Hamiltonian parameters (drive prefactor and bias, detunings, coherent couplings and two-mode squeezing rates) are fixed to arbitrary values and are not optimized. Only the classical linear readout weights are trained.
In contrast, in the trained bosonic QNN, the Hamiltonian parameters are optimized during training. As a consequence, the number of observables required at inference is reduced to a single one, namely the probability of having zero photons in the first mode, $P_1(0)$, compared to nine observables in the reservoir configuration. This leads to a twofold reduction in measurement requirements at inference: (1) fewer observables must be estimated, and (2) since $P_k(0) > P_k(n>0)$, fewer measurement shots are required to reach a fixed statistical precision~\cite{khanPhysicalReservoirComputing2021}. The trained bosonic QNN reaches 100 \% accuracy with 1000 measurement shots (Figure~\ref{fig:fig_sin}(b)), whereas the reservoir configuration of the same hardware required $10^8$ shots~\cite{dudas_quantum_2023}.

This reduction concerns inference only. Training the bosonic QNN requires repeated measurements to estimate gradients over multiple epochs. However, in contrast to reservoir approaches that require estimating several observables, training here involves measuring a single Fock-state probability. In the example considered (Figure~\ref{fig:fig_sin}(b)), the trained network achieves high accuracy with $10^3$ shots per epoch over approximately $10^2$ epochs, corresponding to a total measurement cost below $10^5$ shots. This remains several orders of magnitude smaller than the $\sim 10^8$ shots required to estimate multiple observables in quantum reservoir computing. Therefore, even when training is included, the overall measurement cost is significantly reduced. This cost could be further decreased by pretraining a model of the coupled modes and using it to initialize the physical training, followed by a final fine-tuning stage on the experimental platform to account for model–hardware discrepancies.

\begin{table}
    \centering
    \begin{center}
    \begin{tabular}{l|l|l}
    \hline \hline
                                & \begin{tabular}[c]{@{}l@{}}Quantum\\ reservoir~\cite{dudas_quantum_2023}\end{tabular} & \begin{tabular}[c]{@{}l@{}}Bosonic\\ QNN\end{tabular} \\ \hline
    number of measured states   & 9                                                                                                     & 1                                                     \\ \hline
    number of measurement shots & $10^8$                                                                                                & $10^3$                                                \\ \hline \hline
    \end{tabular}
    \end{center}
    \caption{Number of observables and measurement shots required to reach 100\% accuracy on the sine/square classification task for quantum reservoir and for bosonic QNN. }
    \label{tab:sinsquare}
\end{table}

\subsection*{Encoding schemes}

\begin{figure}[h!]
    \centering\includegraphics[width=0.45\textwidth]{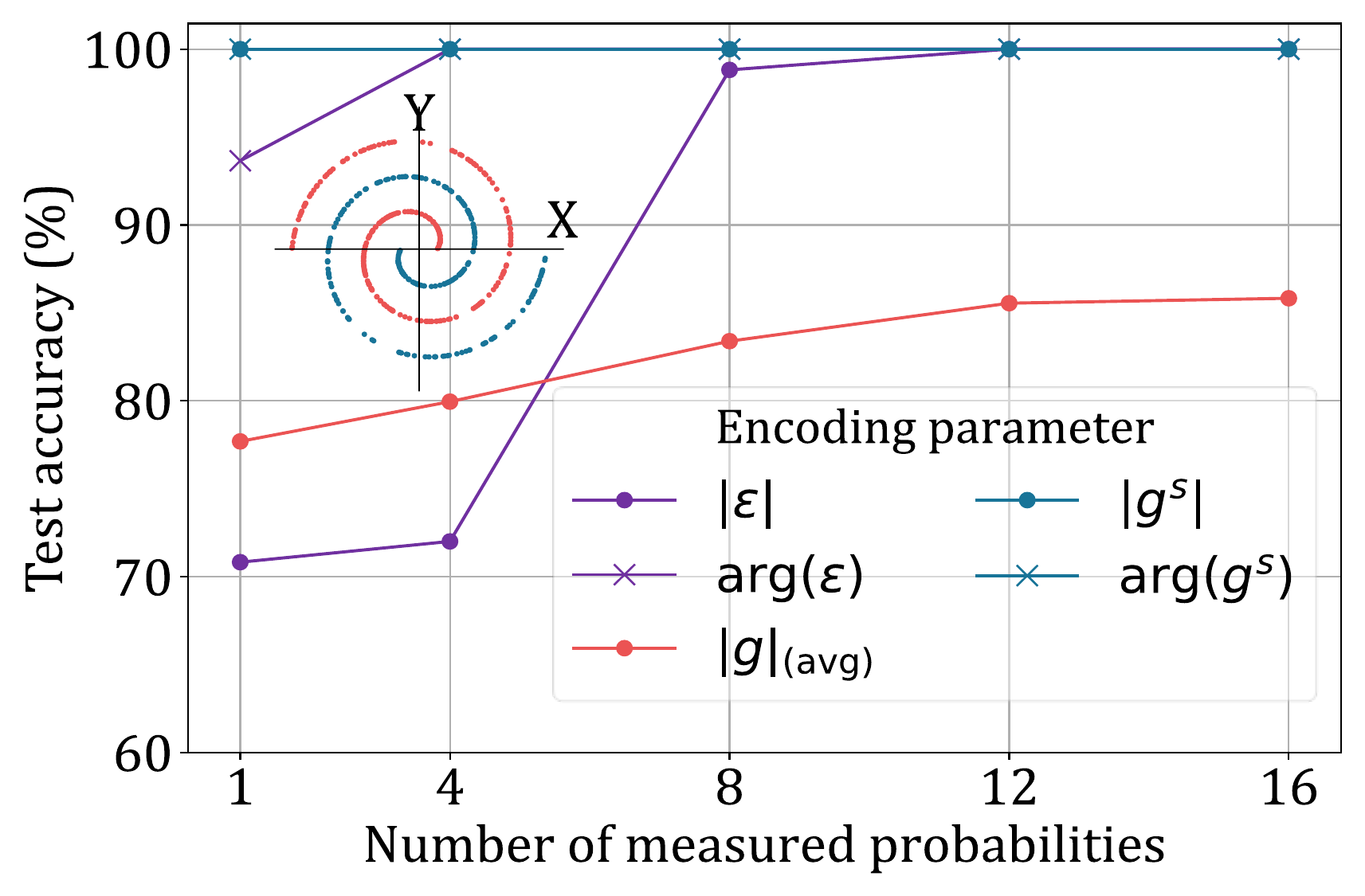}
    \caption{ \textbf{Impact of the encoding variable on the performance on the spirals classification task.} The task consists in assigning a point to the blue or red spiral. The non-linearity of the task is apparent in that it is impossible to draw a straight line to separate the two spirals. The accuracy for different encoding schemes is shown as a function of the number of measured probabilities. Since the accuracies obtained when encoding in $\bf{g}$ vary greatly with the initial physical parameters $\left\{ \bf{g}; \bf{g^s}; \bf{\delta}; \bf{\epsilon}\right\}$, we average them in this case over 5 random initial sets of parameters.}
    \label{fig_spiral}
\end{figure}

Another advantage of analog quantum neural networks is that the choice of the encoding parameter influences the nonlinear transformation that the quantum system applies to the input data. We investigate the optimal encoding using the spirals classification task illustrated in the inset of the Figure~\ref{fig_spiral}. The input data for this task is two-dimensional, it consists in sending the two coordinates of a data point to the network, and labeling it as belonging the the blue or red spiral. This task is harder and requires more nonlinearity than the sine/square classification, we thus address it using four coupled modes. We compare five different encoding schemes: (1) the amplitudes of the nearly resonant drives, (2) their phases, (3) the amplitudes of the two-mode squeezing rates, (4) the phases of the two-mode squeezing rates, and (5) the amplitudes of the exchange coupling rates. For each encoding choice, the selected parameter family carries the input dependence through Eq.~\ref{eq:encoding}, while all other Hamiltonian parameters are treated as trainable. Within the encoding family itself, the global prefactor and bias are optimized too.

For this task, we use the Binary Cross Entropy (BCE) with logits loss function. As shown in Figure~\ref{fig_spiral}, we find that encoding into either the amplitude or phase of the two-mode squeezing rates achieves 100 \% accuracy on the test dataset, with a single measured probability. In contrast, encoding into the drive phase requires 4 measured probabilities to reach 100 \% accuracy, encoding into the drive amplitudes requires 12 measured probabilities, and encoding into the exchange coupling rate amplitude plateaus at 85 \% accuracy. This can be understood by noticing that two-mode squeezing has a more significant impact on the covariance matrix than the coherent photon exchange (see Supp. Mat.). In particular, if there is no two-mode squeezing, the covariance matrix $\bf{\sigma}(t)$ does not evolve beyond its initial vacuum value $\bf{\sigma}_0 = \mathds{1}/2$, independently of the values $\{\bf{\delta}, \bf{\epsilon}, \bf{g}\}$. More generally, while nonlinearity in our model originates from Fock-state measurements, encoding the input data into Gaussian parameters such as the two-mode squeezing rates enhances the effective nonlinear mapping between inputs and measured probabilities. This becomes particularly relevant for highly nonlinear tasks such as spiral classification. A related mechanism has been identified in fully linear optical networks, where nonlinear computation can be achieved by encoding data into the system parameters rather than into the input signal~\cite{wanjura_fully_2024}.

In order to pin down the advantage brought by training the quantum system parameters,  we compare the resources required to reach 100 \% accuracy on the spirals task, in terms of (i) the number of trainable parameters and (ii) the number of observables that must be measured at inference, for both the quantum reservoir configuration and the trained bosonic QNN. As in the previous task, both approaches use the same hardware and the same input encoding; in the reservoir configuration all Hamiltonian parameters are fixed and only the linear readout weights are trained, whereas in the bosonic QNN the Hamiltonian parameters are optimized. The results are summarized in Table~\ref{tab:spiral}. We observe that the bosonic QNN requires a significantly smaller number of measured observables compared to quantum reservoir computing.

As a purely illustrative point of reference, we note that a classical Multi-Layer Perceptron with two hidden layers of six neurons each and 78 trained parameters can also solve this task with comparable accuracy. We stress that this comparison is not intended as a performance benchmark against classical machine-learning architectures, but rather to provide an order-of-magnitude indication of model size.

\begin{table} [h!]
\centering
    \begin{tabular}{l|l|l}
    \hline \hline
                              & \begin{tabular}[c]{@{}l@{}}Quantum\\ Reservoir\end{tabular} & \begin{tabular}[c]{@{}l@{}}Bosonic\\ QNN\end{tabular} \\ \hline
    number of modes $M$       & 4                                                           & 4                                                     \\ \hline
    number of measured states & 36                                                          & 1                                                     \\ \hline
    parameters                & 37                                                          & 38                                                    \\ \hline \hline
    \end{tabular}

    \caption{Number of neurons and parameters needed to reach 100\% accuracy on the spirals classification task using quantum reservoir and bosonic QNN.}
    \label{tab:spiral}
\end{table}

\subsection*{Dynamical stability considerations}

An important practical constraint of our approach is the dynamical stability of driven bosonic systems subject to multiple simultaneous parametric interactions~\cite{bengtsson2018, wintersperger_parametric_2020}. In particular, when several two-mode squeezing processes are applied concurrently, they may interfere constructively and lead to runaway photon generation that cannot be compensated by dissipation or coherent conversion. This photon-number divergence is a fundamental instability of parametric bosonic networks and is not fully mitigated by the regularization term Eq.~\eqref{regularization} alone. To ensure stable training, we therefore explicitly constrain the instantaneous strength of the two-mode squeezing rates during optimization by imposing upper bounds relative to the coherent coupling amplitudes. In practice, this is implemented through a simple clamping procedure that enforces these bounds throughout training. In particular, we impose a maximum amplitude on each two-mode squeezing rate: for phase encoding in the two-mode squeezing rates, the amplitude is limited to $\min(\bf{g})/(M-1)$, where $M$ is the number of modes; for all other encoding schemes, it is limited to half of the smallest amplitude among the coherent coupling rates (see Section VI 
in the Supplementary Material). While heuristic, this stabilization acts only on the training dynamics and does not restrict the set of accessible input–output mappings once a stable operating regime is reached. The precise numerical bounds depend on the chosen encoding scheme and coupling hierarchy, but the need for such constraints is generic and independent of the specific learning task.

Furthermore, stable gradient-based optimization also requires that the linearized dynamical matrix governing the Gaussian evolution of the first and second moments be non-degenerate \cite{KwakernaakSivan1972}. In particular, the evaluation of gradients relies on the diagonalizability of this matrix and on the absence of eigenvalue degeneracies that would lead to ill-conditioned propagators (see Methods, where this matrix is defined explicitly). For this reason, the physical parameters are initialized away from such degenerate points in parameter space, a standard requirement in gradient-based optimization of continuous-variable dynamical systems. During training, the optimization remains within this stable manifold, and we do not observe convergence issues associated with eigenvalue coalescence. This constraint affects the choice of initialization but does not restrict the expressive power of the model.

\subsection*{Handwritten digits classification}

\begin{figure}[h!]
    \centering
    \includegraphics[width=0.5\textwidth]{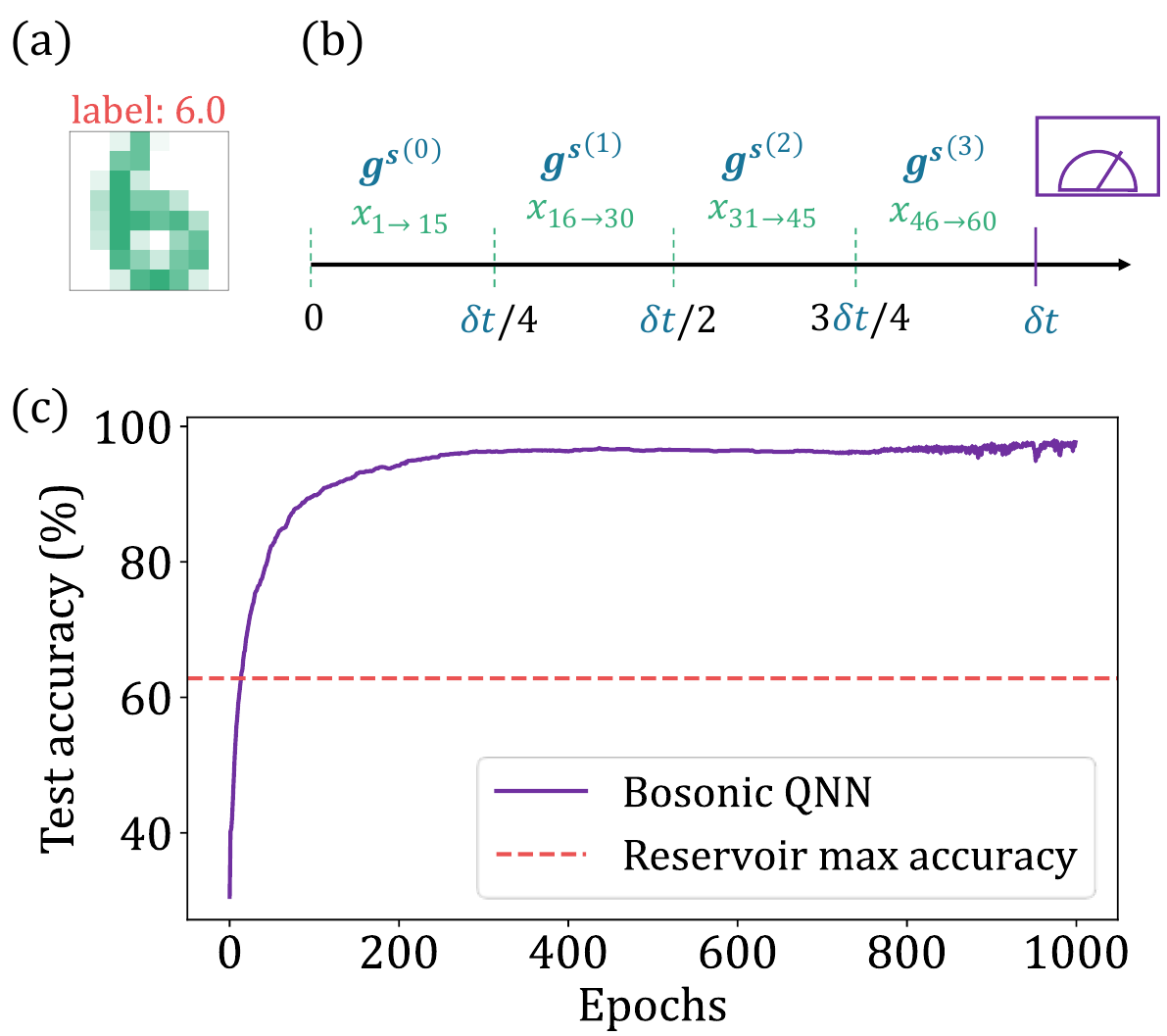}    
    \caption{ \textbf{Handwritten digits classification.} (a) A sample from the DIGITS dataset, consisting of an $8 \times 8$ pixel image and its corresponding label. To create a flattened image vector of size 60, we crop the 4 white corners of the original image. (b) The encoding scheme uses 6 modes. At times $0$, $\delta t/4$, $\delta t/2$, and $3\delta t/4$, 15 pixels are encoded into the amplitudes of the 15 two-mode squeezing rates. During each data re-uploading instance, a new set of parameters $ \{ \bf{g}, \bf{g^s}, \bf{\delta}, \bf{\epsilon} \} $ is applied. At time $\delta t$, the Fock state probabilities are measured, yielding the feature vector $\boldsymbol{F(X)}$. (c) Test set accuracies of the bosonic QNN and quantum reservoir computing with 6 modes. 12 probabilities are measured for the bosonic QNN, while 36 are measured for the quantum reservoir, whose accuracy reaches at best 62.8 \%.}
    \label{fig:digits}
\end{figure}

Training enlarges the set of accessible representations compared to an untrained reservoir and enables the solution of tasks that are otherwise inaccessible when the same hardware is used in a reservoir-computing configuration. We demonstrate this by solving the handwritten digit recognition task from the scikit-learn DIGITS dataset, shown in Figure~\ref{fig:digits}a.
We use 6 modes, pairwise coupled through 15 two-mode squeezing processes. The amplitudes of these squeezing terms are used for data encoding. Since 64-pixel images cannot be processed in a single time step, we adopt an encoding scheme inspired by the data re-uploading method~\cite{perez-salinasDataReuploadingUniversal2020}: after removing the 4 white corner pixels, each image is divided into four 15-pixel batches, which are sequentially injected over four time intervals $\frac{\delta t}{4} = \frac{2\pi}{10\kappa}$, as shown in Figure~4(b).
In the trained bosonic QNN, the Hamiltonian parameters (including the squeezing prefactor and bias, drive amplitudes, detunings and coherent couplings) are optimized jointly with the classical output weights. In the reservoir configuration, the same 6-mode architecture, encoding protocol, and time evolution are used, but all Hamiltonian parameters are fixed to random values and only the linear readout weights are trained.
The scikit-learn DIGITS dataset contains 1797 samples; for computational convenience, we use a randomly selected subset of 1500 samples, which is sufficient for this study. Using the trained configuration, we achieve over 97 \% accuracy by measuring 12 probability amplitudes $P_k(n)$ and training a total of 502 Hamiltonian and output parameters. In contrast, the reservoir configuration with identical hardware achieves at best 62.8 \% test accuracy, even when measuring 36 probability amplitudes (Table~\ref{tab:digits}).

\begin{table} [h!]
\centering
    \begin{tabular}{l|l|l}
    \hline \hline
  & \begin{tabular}[c]{@{}l@{}}Quantum\\ Reservoir\end{tabular} & \begin{tabular}[c]{@{}l@{}}Bosonic\\ QNN\end{tabular} \\ \hline
number of modes $M$ & 6  & 6  \\ \hline
number of measured states & 36 & 12   \\ \hline
parameters          &  370   &  502 \\ \hline
accuracy               &  62.8 \% & 97.1 \%  \\ \hline \hline
    \end{tabular}

    \caption{ Comparison between quantum reservoir computing and a trained bosonic QNN in terms of number of modes, measured Fock-state probabilities, trainable physical parameters, and achieved test accuracy.}
    \label{tab:digits}
\end{table}

\section*{Discussion}

We have introduced a training framework for bosonic quantum neural networks based on parametrically driven Gaussian dynamics that can be efficiently simulated and differentiated in the Heisenberg representation. Because the underlying evolution is linear in the quadrature operators and governed by quadratic Hamiltonians, gradients with respect to physical parameters can be computed via backpropagation through a classical model of the system. Nonlinearity, which is essential for expressive learning, arises from two complementary mechanisms: Fock-basis measurements at the output and the encoding of input data directly into Gaussian parameters, in particular the two-mode squeezing rates. Their combination yields a flexible and highly nonlinear input–output map while retaining a tractable forward model.
Optimizing the physical parameters enables the network to solve increasingly complex tasks without enlarging the underlying hardware. At the same time, training substantially reduces the measurement resources required at inference. For example, in the sine/square and spiral classification tasks, a single measured probability is sufficient after training, compared to 9 and 36 observables, respectively, in reservoir-based implementations of similar size. This reduction makes experimental inference significantly more practical once parameters have been optimized. Even when accounting for the measurements required during training, the total number of shots in our examples remains well below that required to estimate multiple observables in untrained reservoirs.
We further find that encoding data into two-mode squeezing parameters is particularly advantageous compared to encoding in coherent drives or exchange couplings. Physically, two-mode squeezing directly reshapes the covariance matrix and therefore enhances the curvature of the mapping from encoded parameters to measured Fock probabilities. This additional source of effective nonlinearity becomes especially beneficial for highly nonlinear tasks such as spiral classification. More generally, our results highlight that expressive nonlinear computation can be achieved in systems with linear Gaussian dynamics when nonlinearity is introduced through measurement and parameter encoding.

A limitation of the present approach is that gradient-based training relies on classical simulation of the Gaussian dynamics, which ultimately restricts scalability. When moving toward larger experimental implementations, the choice of optimization strategy will be central in determining the overall resource cost. In particular, gradient-free or hybrid optimization methods may provide a practical alternative and would enable the exploration of extended parameter spaces, including higher-order mixing processes such as four-wave interactions.
Finally, the training properties of bosonic quantum neural networks remain largely unexplored. In parametrized quantum circuits, barren plateaus can hinder gradient-based optimization under certain conditions~\cite{larocca_barren_2025}. Whether analogous gradient suppression phenomena arise in continuous-variable, driven-dissipative Gaussian architectures is not yet understood. While we do not address this question systematically here, our results demonstrate that stable gradient-based training is feasible for moderately sized bosonic networks under physically motivated constraints. A detailed investigation of optimization landscapes, scalability limits, and potential gradient concentration effects in bosonic architectures constitutes an important direction for future work.

\section*{Methods}

\subsection*{Solution of the quantum Langevin equation}
We have performed the simulations using the PyTorch Python library, and in particular its automatic implementation of the backpropagation algorithm. To write the time evolution of the displacement vector $\bf{\alpha}(t)$ and covariance matrix $\bf{\sigma}(t)$ which fully determine the Gaussian state, we express the field operators $\hat{a}_k(t)$ in the Heisenberg representation, where they depend on time and follow the quantum Langevin equation
\begin{equation}
	\label{eq:Langevin}
	\frac{d \hat{a}_k }{dt} = -\frac{i}{\hbar} [\hat{a}_k, \hat{H}_0] -\frac{\kappa_{k}}{2} \hat{a} - \sqrt{\kappa_k} \hat{a}_{k, \text{in}}.
\end{equation}
In this picture, the complex drive which was modeled through the drive Hamiltonian $\hat{H}_{\text{in}}$ in Eq.~\eqref{eq:Hamiltonian}, intervenes through the $ \sqrt{\kappa_k} \hat{a}_{k, \text{in}}$ term. The first and second moments $\bf{\alpha}$ and $\bf{\sigma}$ of the field operator $\hat{a}_k(t)$ are defined as
{
\footnotesize
\begin{equation}
\label{eq:alpha_sigma_def}
    \begin{cases}
        \hat{A}(t) & = ( \hat{a}_1(t), \ldots, \hat{a}_M(t),
        \hat{a}_1^\dagger (t), \ldots, \hat{a}_M^\dagger (t))^T \\
        \bf{\alpha}(t) & = \langle \hat{A}(t) \rangle  \\
        \bf{\sigma}_{k,l}(t) & = \frac{1}{2} \left[ \langle \hat{A}_k(t) \hat{A}_l(t)^\dagger \rangle + \langle \hat{A}_l(t)^\dagger \hat{A}_k(t) \rangle \right] - \bf{\alpha}_k(t) \bf{\alpha}_l^*(t). 
    \end{cases}
\end{equation}
}
The vectorized Langevin equation for the entire system is
\begin{equation}
    \label{eq:Langevin_A}
    \frac{d\hat{A}}{dt}  = \mathcal{L}\hat{A} - \frac{K}{2}\hat{A} - \sqrt{K}\hat{A}_{\text{in}}, 
\end{equation}
where $K = \text{diag}(\kappa_{1}, \cdots, \kappa_M , \kappa_{1}, \cdots, \kappa_M)$ and $\hat{A}_{\text{in}} = \left( \hat{a}_{1, \text{in}},  \cdots, \hat{a}_{M, \text{in}}, \hat{a}_{1, \text{in}}^\dagger, \cdots \right)^T$. $\mathcal{L}$ is a coupling matrix, expressed as
\begin{equation}
	\label{eq:L_formula}
	\mathcal{L} = \frac{1}{i\hbar}
	\begin{pmatrix}
		G & G^s \\
		-{G^s}^\dagger & -G^T
	\end{pmatrix},
\end{equation}
where the matrix elements are
{
\footnotesize
\begin{equation*}
	(G)_{k,l} = \hbar \times 
	\begin{cases}
		-\delta_k & \text{if } k = l \\
		g_{k,l} & \text{if } k < l \\
		g^*_{k,l} & \text{if } k > l
	\end{cases}, \quad
	(G^s)_{k,l} = \hbar \times 
	\begin{cases}
		0 & \text{if } k = l \\
		{g^s}_{k,l} & \text{otherwise.}
	\end{cases}
\end{equation*}
}
This differential equation has the following solution~\cite{gouzienOptiqueQuantiqueMultimode2019}:
\begin{equation}
\label{eq:A_t_integral}
    \hat{A}(t) = F(t) \hat{A}(t=0) - \int_0^t F(t-\tau) \sqrt{K} \hat{A}_{\text{in}}(\tau) d\tau,
\end{equation}
where we define the propagator matrix
\begin{equation}
    F(t) = \exp(F' t), \quad \text{with } F' = \mathcal{L} - \frac{K}{2}.
\end{equation}

\subsection*{Computation of the displacement and covariance matrix of field operators via diagonalization}

We compute the displacement $\bf{\alpha}(t)$ and covariance matrix $\bf{\sigma}(t)$ of the field operators by using Eq.~\eqref{eq:A_t_integral} and their definitions in Eq.~\eqref{eq:alpha_sigma_def}:
{
\footnotesize
\begin{equation}
    \label{eq:alpha_sigma_t}
    \begin{cases}
        \bf{\alpha}(t) & = F(t) \bf{\alpha}(t=0) - \int_0^t F(t-\tau) \sqrt{K} d\tau \bf{\alpha}_{\text{in}}\\
        \bf{\sigma}(t) & = F(t)\bf{\sigma}(t=0)F^\dagger(t)+\bf{\sigma}_{0} \int_0^t F(t-\tau) K F^\dagger(t-\tau) d\tau,
    \end{cases}
\end{equation}
}
where $\bf{\sigma}_0 = \frac{1}{2} \mathds{1}_M$ is the vacuum covariance, and $\bf{\alpha}_{\text{in}} = ( \epsilon_1, \hdots, \epsilon_M, \epsilon_1^*, \hdots, \epsilon_M^* )^T$ the input coherent drive. We set the inputs modes $\hat{A}_{\text{in}}$ to be in coherent states of constant values $\bf{\alpha}_{\text{in}}$. The calculation of $\bf{\sigma}(t)$ is provided in section VIII of the Supplementary Material. Assuming that $F'$ is diagonalizable as $F' = U \Lambda U^{-1}$ with $\Lambda = \mathrm{diag}(\lambda_1, \ldots, \lambda_{2M})$, the matrix exponential becomes:
\begin{equation}
    F(t) = U e^{t \Lambda} U^{-1}.
\end{equation}
Then, the integral of $\bf{\alpha}(t)$ in Eq.~\eqref{eq:alpha_sigma_t} becomes
\begin{equation}
\label{eq:alpha_calculation}
    \begin{split}
        \bf{\alpha}(t) & = F(t)\mathbf{\alpha}(t=0) - \sqrt{K} U \int_0^t e^{\Lambda (t-\tau)} d\tau U^{-1} \mathbf{\alpha}_{\text{in}}\\
        & = F(t)\mathbf{\alpha}(t=0) - \sqrt{K} U I_1 U^{-1} \mathbf{\alpha}_{\text{in}},
    \end{split}
\end{equation}
where $I_1 = \Lambda^{-1} (e^{\Lambda t} - \mathds{1}_{2M})$. To compute the covariance matrix $\bf{\sigma}(t)$, we introduce the matrices $P$ and $I_2$ such that
\begin{equation}
    \begin{cases}
        P & = U^{-1} K (U^{-1})^\dagger \\
        (I_2)_{i,j} & = (P)_{i,j} \frac{e^{(\lambda_i + \lambda_j^*)t} - 1}{\lambda_i + \lambda_j^*}.
    \end{cases}
\end{equation}
Finally, we find
\begin{equation}
    \label{eq:sigma_calculation}
    \begin{split}
        \bf{\sigma}(t) = F(t)\bf{\sigma}(t=0)F(t)^{\dagger} + \bf{\sigma}_0 U I_2 U^\dagger.
    \end{split}
\end{equation}

\subsection*{Gaussian Boson Sampling} \label{sec:GBS}

To compute the joint probability of obtaining $n_k$ photons in each mode $k$ from the displacement and covariance matrix $\{\bf{\alpha}, \bf{\sigma}\}$, we first define some intermediary variables
\begin{equation}
\label{eq:GBS_intermediate_matrices}
    \begin{cases}
        \bf{\sigma}_{Q} & = \bf{\sigma} + \mathds{1}_{2M} / 2\\
        \bf{T} & = \begin{pmatrix}
            0_{M} & \mathds{1}_{M} \\ \mathds{1}_{M} & 0_{M}
        \end{pmatrix}\\
        \bf{A} & = \bf{T} \left( \mathds{1}_{2M} - \bf{\sigma}_{Q}^{-1} \right) \\
        \bf{\gamma} & = \bf{\alpha}^\dagger \bf{\sigma}_Q^{-1}.
    \end{cases}
\end{equation}
Given the photon number vector $\bar{n}=(n_k)_{k\in [1,M]}$, we construct $\bf{A}_{\bar{n}}$ from $\bf{A}$ by repeating $k$th column and rows $n_k$ times. Similarly, $\bf{\gamma}_{\bar{n}}$ is constructed from $\bf{\gamma}$ by repeating $k$th column and rows $n_k$ times. Then the diagonal elements of $\bf{A}_{\bar{n}}$ are substituted by $\bf{\gamma}_{\bar{n}}$.
Using the variables of Eq.~\eqref{eq:GBS_intermediate_matrices}, the GBS formula gives the joint Fock state probability of measuring the photon combination $\bar{n}$:
\begin{equation}
    P(\bar{n}, \bf{\alpha}, \bf{\sigma}) =  \frac{\exp(-\frac{1}{2} \bf{\alpha}^\dagger \bf{\sigma}_Q^{-1} \bf{\alpha})}{\sqrt{\mathrm{det}(\bf{\sigma}_Q)} \prod_k n_k!} \mathrm{lhaf}(\bf{A}_{\bar{n}\oplus\bar{n}}),
\end{equation}
where $\bar{n}\oplus\bar{n}$ is $\bar{n}$ concatenated with itself, so that $\bf{A}_{\bar{n}\oplus\bar{n}}$ is constructed from $\bf{A}$ by repeating $k$th and $(k+M)$th column and rows $n_k$ times, and replacing its diagonal by $\bf{\gamma}_{\bar{n}\oplus\bar{n}}$.

The field operator moments of the partially traced Gaussian state over all modes except $k$ are obtained by only keeping the $k$th and $k+M$th columns and rows in $\bf{\alpha}$ and $\bf{\sigma}$. We denote these partially traced moments $\bf{\alpha}_k$ and $\bf{\sigma}_k$. Applying the GBS formula to them recovers the local Fock state probability of Eq.~\eqref{eq:GBS}

\subsection*{Benchmark tasks}
\label{sec:benchmark_tasks}

 Input data for all the benchmark tasks are rescaled to lie within the interval $[0, 1]$. This normalization facilitates more controlled tuning of the encoding parameters $\bf{\theta}$, ensuring that their absolute values do not exceed $|\bf{\theta}_0 + \bf{\theta}_{\text{bias}}|$, as defined in Eq. (2). Initial parameters and hyper-parameters used for each task are listed in Table~\ref{tab:task_params}.

\begin{table}[h!]
    \centering
    \begin{subtable}{0.7 \textwidth}
    \begin{minipage}[t!]{0.62\textwidth}
    \centering
    \begin{tabular}{@{}lll@{}}
    \toprule
    parameter              & initial value range             & learning rate \\ \midrule
    $\bf{W}_0$             & 1                               & 0.01          \\
    $\bf{W}_{\text{bias}}$ & 0                               & 0.01          \\
    $\kappa$               & $2\pi \times$ 2 MHz                   & none          \\
    $\bf{\kappa}$          & $(1 \pm 0.1)\kappa $            & none          \\
    $\bf{\delta}$          & 0                    & 0.1           \\
    $\bf{\epsilon}$        & $ (170 \pm 30) \sqrt{\kappa/4\pi}$ & 0.1           \\
    $\bf{g}$               & $45\kappa$                      & 0.1           \\
    $\bf{g^s}$             & $9\kappa$                       & 0.1           \\
    $\delta t$             & $0.4 \pi \kappa^{-1}$                    & none          \\ \bottomrule
    \end{tabular}
   \end{minipage}
    \hfill
    \begin{minipage}[t!]{0.34\textwidth}
    \centering
    \begin{tabular}{@{}ll@{}}
    \toprule
    Hyper-parameter                        & Value        \\ \midrule
    modes                                  & 2            \\
    epochs                                 & 500          \\
    $\beta_{\langle\bar{N}\rangle}$        & 0.02         \\
    $\langle \bar{N}\rangle_{\text{thr}} $ & 3            \\
    $\langle \bar{N}\rangle_{\text{tg}}$   & 2            \\
    batches                                & 5            \\
    dataset size                           & 200          \\
    loss                                   & $\text{MSE}$ \\ \bottomrule
    \end{tabular}
    \end{minipage}

    \subcaption{Sine/square classification task learning parameters}
    \label{subtab:sinsquare_params}
    \end{subtable}

    \vspace{1em}

    \begin{subtable}{0.7\textwidth}
    \begin{tabular}{@{}lll@{}}
    \toprule
    parameter              & initial value range     & learning rate \\ \midrule
    $\bf{W}_0$             & 1                       & 0.1           \\
    $\bf{W}_{\text{bias}}$ & 0                       & 0.1           \\
    $\kappa$               & $2\pi \times$ 2 MHz           & none          \\
    $\bf{\kappa}$          & $(1 \pm 0.1)\kappa $    & none          \\
    $\bf{\delta}$          & $(0.5 \pm 0.01)\kappa$  & 0.1           \\
    $\bf{\epsilon}$        & $ 400 \sqrt{\kappa/4 \pi} $ & 0.1           \\
    $\bf{g}$               & $(50 \pm 5)\kappa$      & 0.1           \\
    $\bf{g^s}$             & $(10 \pm 1)\kappa$      & 0.1           \\
    $\delta t$             & $0.8 \pi \kappa^{-1}$            & none          \\ \bottomrule
    \end{tabular}
    \qquad
    \begin{tabular}{@{}ll@{}}
    \toprule
    Hyper-parameter                        & Value                    \\ \midrule
    modes                                  & 4                        \\
    epochs                                 & 500                      \\
    $\beta_{\langle\bar{N}\rangle}$        & 0.02                     \\
    $\langle \bar{N}\rangle_{\text{thr}} $ & 3                        \\
    $\langle \bar{N}\rangle_{\text{tg}}$   & 2                        \\
    batches                                & 5                        \\
    dataset size                           & 500                      \\
    loss                                   & $\text{BCE with logits}$ \\ \bottomrule
    \end{tabular}
    \subcaption{Spiral classification task learning parameters}
    \label{subtab:spiral_params}
    \end{subtable}

    \vspace{1em}
    
    \begin{subtable}{0.7\textwidth}
    \begin{tabular}{@{}lll@{}}
    \toprule
    parameter              & initial value range   & learning rate \\ \midrule
    $\bf{W}_0$             & 1                     & 0.01          \\
    $\bf{W}_{\text{bias}}$ & 0                     & 0.01          \\
    $\kappa$               & $2 \pi \times $ 2 MHz         & none          \\
    $\bf{\kappa}$          & $(1 \pm 0.1)\kappa $  & none          \\
    $\bf{\delta}$          & 0          & 0.01          \\
    $\bf{\epsilon}$        & $ 600\sqrt{\kappa/4 \pi}$ & 0.01          \\
    $\bf{g}$               & $(50 \pm 5)\kappa$    & 0.01          \\
    $\bf{g^s}$             & $(10 \pm 1)\kappa$    & 0.01          \\
    $\delta t$             & $0.8 \pi \kappa^{-1}$          & none          \\ \bottomrule
    \end{tabular}
    \qquad
    \begin{tabular}{@{}ll@{}}
    \toprule
    hyper-parameter                        & value                  \\ \midrule
    modes                                  & 6                      \\
    epochs                                 & 1000                   \\
    $\beta_{\langle\bar{N}\rangle}$        & 0.12                   \\
    $\langle \bar{N}\rangle_{\text{thr}} $ & 3                      \\
    $\langle \bar{N}\rangle_{\text{tg}}$   & 2                      \\
    batches                                & 5                      \\
    dataset size                           & 1500                   \\
    loss                                   & $\text{Cross Entropy}$ \\ \bottomrule
    \end{tabular}
    \subcaption{DIGITS classification task learning parameters}
    \label{subtab:digits_params}
    \end{subtable}

    \caption{Learning parameters for the (a) sine/square, (b) spirals, and (c) DIGITS classification tasks. $\kappa$ is the average value of the dissipation rates $\bf{\kappa}$. For all the tasks, the dataset sizes specified are the same for the training and the testing sets.}
    \label{tab:task_params}
\end{table}

\subsubsection*{Sine and square waveform classification}

The physical features vector $\bf{F(X)}$ includes a single component $P_1(0)$. After training, the average photon number corresponding to the maximum input value is found to be $\langle\bar{N}\rangle = 8$.

\subsubsection*{Spirals classification task}

This task uses a two-class spirals dataset generated from points in polar coordinates according to
\begin{equation}
    \begin{cases}
        \theta(\xi) & \sim \mathcal{U}(0, 3\pi), \\
        r(\xi) & = \pm \dfrac{2\theta(\xi) + \pi}{25},
    \end{cases}
\end{equation}
where $\mathcal{U}(a, b)$ denotes the uniform distribution on the interval $[a, b]$. Points with a positive (negative) sign in $r(\xi)$ are labeled as class 1 (class 0).  The input data is symmetric with respect to the origin in the 2-dimensional input plane. To incorporate this symmetry into the model, we augment each input point $[x_0, x_1]$ to $[x_0, x_1, -x_0, -x_1]$. Each input vector $\mathbf{x}$ of dimension $s_{\mathbf{x}}$ can be encoded in the phase of the encoding parameter $\bf{\theta}$, resulting in the modified encoding:
\begin{equation}
\label{eq:encoding_phase}
\begin{split}
    \bf{\theta}(\mathbf{x}) &= \bf{\theta}_0 e^{i \bf{\varphi}(\mathbf{x})} + \bf{\theta}_{\text{bias}}, \\
    \bf{\varphi}(\mathbf{x}) &= \bf{\varphi}_0 \mathbf{x} + \bf{\varphi}_{\text{bias}},
\end{split}
\end{equation}
where $\bf{\theta}_0, \bf{\theta}_{\text{bias}} \in \mathbb{C}^{s_{\mathbf{x}}}$, and $\bf{\varphi}_0, \bf{\varphi}_{\text{bias}} \in \mathbb{R}^{s_{\mathbf{x}}}$. We initialize the phase parameters as $(\bf{\varphi}_0)_i = \pi$ and $(\bf{\varphi}_{\text{bias}})_i = 0$ for all $i \in \{1, \dots, s_{\mathbf{x}}\}$. As in the previous task, the measurement consists of a single probability, $P_1(0)$. After training, the average photon number for the maximum input is $\langle\bar{N}\rangle = 10$. The Binary Cross Entropy (BCE) with logits loss is implemented using PyTorch. It consists of two steps: applying the element-wise sigmoid function $x \mapsto \frac{1}{1 + e^{-x}}$ to the predictions, followed by the BCE computation:
\begin{equation}
    \text{BCE}(x, y) = y \log(x) + (1 - y) \log(1 - x),
\end{equation}
where $x$ and $y$ denote the prediction and target labels, respectively.

\subsubsection*{DIGITS classification task}

For this task, we use the DIGITS dataset from the \texttt{scikit-learn} Python library. Inputs are encoded in the amplitude of the two-mode squeezing rates $|g^s|$, with initial parameters $\bf{\theta}_0 = \vec{1}$ and $\bf{\theta}_{\text{bias}} = \vec{0}$. Measurements are taken over the probabilities $P_k(n)$ for $k \in \{1, 2, 3, 4\}$ and $n \in \{0, 1, 2\}$, resulting in a feature vector $\bf{F(X)}$ with 12 components. After training, the average photon number for the maximum input is $\langle\bar{N}\rangle = 20$. The Cross Entropy loss is implemented in PyTorch. Predictions are first passed through the softmax function, followed by the computation of the cross-entropy between the predicted class distribution and target label.

\color{black}

\bibliography{Main/Mainbib}

\section*{Ackowledgements}
This research was supported by the European Union (ERC, qDynnet, 101076898). Views and opinions expressed are however those of the authors only and do not necessarily reflect those of the European Union or the European Research Council.
Neither the European Union nor the granting authority can be held responsible for
them.

\section*{Author contributions statement}
D.M. and J.G. conceived the project. J.D. and E.G. performed the calculations, J.D. performed the quantum simulations. B.C. participated to analytical calculations and quantum simulations. D.M. and J.D. wrote the manuscript.

\section*{Additional information}
The code and data that support this study are available in Zenodo with the identifier https://doi.org/10.5281/zenodo.15856611

\end{document}


\title{Supplementary Material for "Training the parametric interactions in an analog bosonic quantum neural network with Fock basis measurement"}

\author{J.~Dudas}
\author{B.~Carles}
\author{J. Grollier}
\affiliation{
Laboratoire Albert Fert, CNRS, Thales, Université Paris-Saclay, 91767 Palaiseau, France
}
\author{E. Gouzien}
\affiliation{Alice \& Bob, 53 Bd du Général Martial Valin, 75015 Paris, France }
\author{D.~Marković}
\affiliation{
Laboratoire Albert Fert, CNRS, Thales, Université Paris-Saclay, 91767 Palaiseau, France
}

\date{\today}

\maketitle

\tableofcontents

\newpage

\section{Influence of the two-mode squeezing and photon conversion rates on the covariance matrix} \label{sec:squeezing_influence_on_cov}

To understand why encoding in the two-mode squeezing rates leads to better performance than encoding in the coherent photon conversion rates, we examine their influence on the covariance matrix $\bf{\sigma}(t)$.

\begin{figure}[h]
    \centering
    \includegraphics[width=0.7\linewidth]{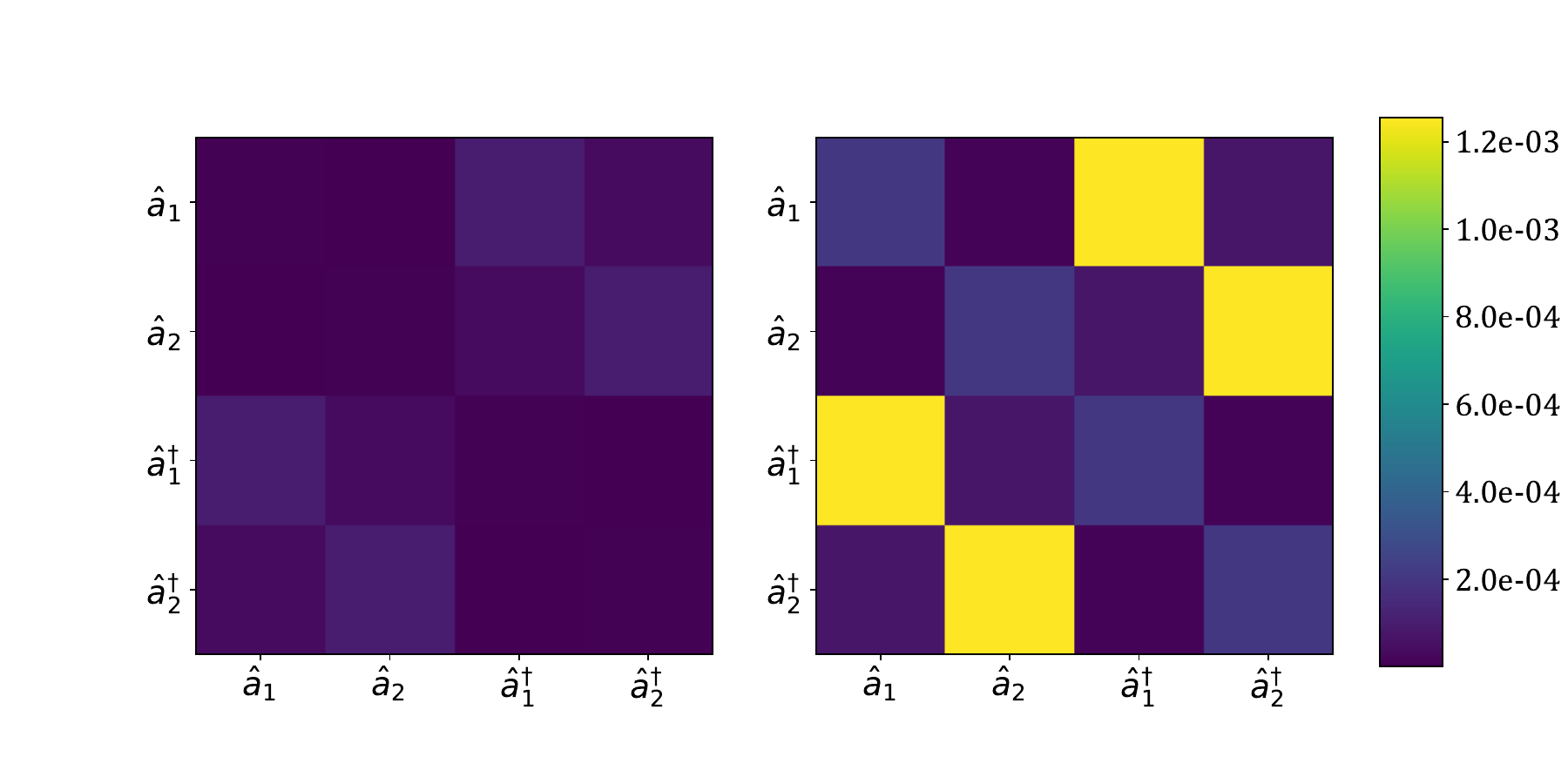}
    \caption{Variance matrix of the ladder operator covariance matrix $|\bf{\sigma}|$ for two modes after $\delta t=0.8 \pi \kappa^{-1}$ of time evolution, with $\kappa = 2 \pi \times \SI{2}{MHz}$. \textbf{Left:} Varying the photon conversion rate $g$ from $45 \kappa$ to $55 \kappa$ with fixed two-mode squeezing rate $g^s = 10 \kappa$. \textbf{Right:} Varying the two-mode squeezing rate $g^s$ from $5 \kappa$ to $15 \kappa$ with fixed $g = 50 \kappa$. The dissipation rates are $\kappa_1 = \kappa_2 = \kappa$.}
    \label{fig:sigma_dep_g_gs}
\end{figure}

First, it is straightforward to show that if $\bf{g^s} = 0$, then for all $t$ the covariance matrix remains constant: $\bf{\sigma}(t) = \bf{\sigma}_0 = \frac{1}{2} \mathbb{1}_{2M}$. Next, we analyze the variance of the covariance matrix as a function of the photon conversion rate $g$ and the two-mode squeezing rate $g^s$ in the case of two coupled modes. As shown in Fig.~\ref{fig:sigma_dep_g_gs}, every term of the covariance matrix exhibits lower variance when $g$ is varied compared to when $g^s$ is varied. 

\section{Comparison to a Multi-Layer Perceptron}
To provide a classical baseline for our bosonic QNN, we compare its performance on the spirals classification task with that of a Multi-Layer Perceptron (MLP). Unlike the DIGITS task, which can be solved by a single-layer perceptron, the spirals task requires at least two hidden layers due to its higher nonlinearity. This is why we chose the spirals task for this comparison.

\begin{figure}[h]
    \centering
    \includegraphics[width=0.5\linewidth]{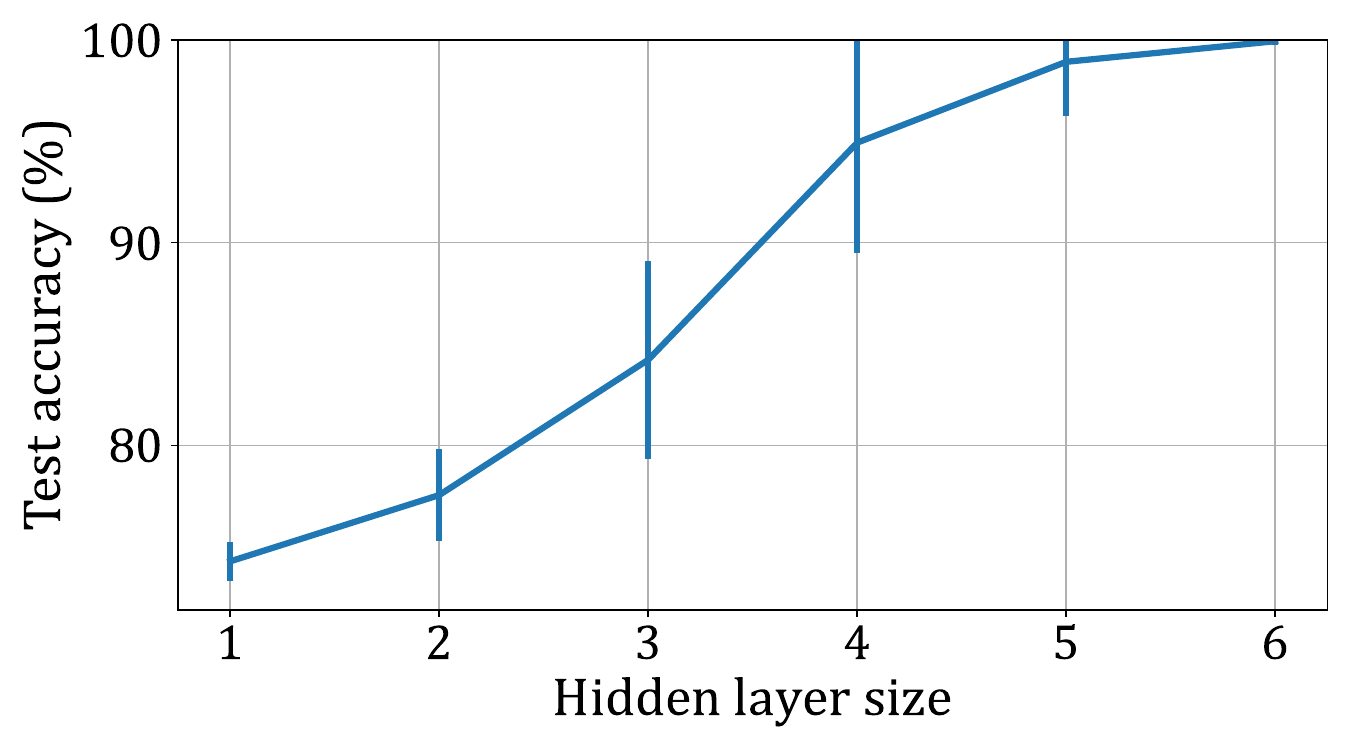}
    \caption{Test accuracy of a 2-layer MLP on the spirals classification task as a function of the number of neurons per hidden layer. The test accuracies are averaged over 19 different random initializations of the MLP weights, and the error bars correspond to the standard deviation.}
    \label{fig:MLP_spiral_perf}
\end{figure}

We use the $\tanh$ activation function, as it outperformed ReLU in this setting. The output layer consists of a single neuron. Letting $n_\text{h}$ denote the number of neurons per hidden layer, the total number of trainable parameters is $5n_\text{h} + (n_\text{h} + 1)^2$. Figure~\ref{fig:MLP_spiral_perf} shows the model's performance as a function of $n_\text{h}$. We find that at least $6$ neurons per hidden layer are required to achieve 100\% classification accuracy, corresponding to $79$ trainable parameters—more than double the $38$ parameters used by the bosonic QNN.

\section{Renormalization of physical parameters}
In optimization algorithms, it is preferable for all learned parameters to be of the same order of magnitude. To achieve this, we apply a rescaling of physical parameters using a renormalization factor $\mathcal{R} = 10^7$ in all simulations:
\begin{equation}
\begin{cases}
    t (s) \rightarrow & t \times \mathcal{R} \\
    \bf{\delta} (\text{Hz}) \rightarrow & \bf{\delta} / \mathcal{R} \\
    \bf{g} (\text{Hz}) \rightarrow & \bf{g} / \mathcal{R} \\
    \bf{g^s} (\text{Hz}) \rightarrow & \bf{g^s} / \mathcal{R} \\
    \bf{\kappa} (\text{Hz}) \rightarrow & \bf{\kappa} / \mathcal{R} \\
    \bf{\epsilon} (\sqrt{\text{Hz}}) \rightarrow & \bf{\epsilon} / \sqrt{\mathcal{R}}.
\end{cases}
\end{equation}

This rescaling corresponds to a change of time units $t (s) = t \times \mathcal{R}$, under which the master equation retains its form provided all frequency-like quantities are scaled as $\omega \rightarrow \omega/ \mathcal{R}$. The drive amplitudes scale as $\epsilon \rightarrow \epsilon / \mathcal{R}$ to preserve the photon-number injection rate. Therefore, this transformation does not modify the physical dynamics but improves numerical conditioning by bringing all trainable parameters to comparable magnitudes.

\section{Derivatives with respect to eigenvectors}
The calculation of the ladder operator displacement $\bs{\alpha}(t)$ and covariance matrix $\bs{\sigma}(t)$ in Eqs. (13, 15) rely on the eigenvectors $U$ of the matrix $F' = \mathcal{L} - \frac{K}{2}$. However, as stated in the PyTorch documentation~\cite{TorchlinalgeigPyTorch262024}, gradients involving eigenvectors of a matrix $A$ are only well defined if $A$ has distinct eigenvalues. Gradients become unstable when eigenvalues are nearly degenerate.

A simple example with $M=2$ modes, identical dissipation rates $\kappa_1 = \kappa_2 = \kappa$ and zero detunings $\delta_1 = \delta_2 = \SI{0}{Hz}$ illustrates this issue. The eigenvalues $(\lambda_-, \lambda_+)$ of $F'$ in this case are two-fold degenerate:
\begin{equation}
\label{eq:F'_eigvals_M=2}
\begin{cases}
    \lambda_\pm = \pm i\sqrt{|g|^2 - |g^s|^2} - \frac{\kappa}{2} &\text{if $|g|>|g^s|$}\\
    \lambda_\pm = \pm \sqrt{|g^s|^2 - |g|^2} - \frac{\kappa}{2} &\text{if $|g^s|>|g|$}\\
\end{cases}.
\end{equation}
In this scenario, gradient computation will fail due to the degeneracy. To avoid such issues, the initial physical parameters $\left\{ \bs{\epsilon}, \bs{\delta}, \bs{g}, \bs{g^s}, \bs{\kappa} \right\}$ are chosen such that $F'$ has non-degenerate eigenvalues.

\section{Divergence of the mean photon number \texorpdfstring{$\langle N \rangle$}{<N>}}
\label{sec:divN}
In order to avoid the divergence of photon number expectation values $\langle N \rangle$, we have introduced a regularization term $\text{loss}_{\langle N \rangle}$ to the loss function, that penalizes high $\langle N \rangle$. However this is not sufficient to prevent abrupt divergences of $\langle N \rangle$, for certain coupling rate $\{ \bs{g}, \bs{g^s} \}$ values. We will introduce different special cases in order to understand where they come from. In all of the examples of this section
\begin{equation}
\label{eq:N_study_default_params}
    \forall i\in [1, M] \begin{cases}
        \kappa_i &= \kappa = 2\pi \times\SI{2}{MHz} \\
        \delta_i &= \SI{0}{Hz} \\
        \epsilon_i &= \epsilon = M\times 100 \times \sqrt{\kappa/4 \pi}
    \end{cases},
\end{equation}
and the initial state is vacuum, such that the initial displacement and covariance matrix are $\bs{\alpha}(t=0) = \Vec{0}$ and $\bs{\sigma}(t=0) = \bs{\sigma}_0 = \frac{1}{2} \mathbb{1}_{2M}$.

\subsection{Computation of the mean photon number \texorpdfstring{$N(t)$}{N(t)} starting from vacuum} \label{sec:N_t_calc}
We assume $F'=\mathcal{L}-\frac{K}{2}$ is diagonalizable into $F'=U\Lambda U^{-1}$, and $\Lambda=\text{diag}(\lambda_1, \dots, \lambda_{2M})$. This lets us compute the integral of Eq.~(9) to get
\begin{equation}
\label{eq:A_t_diag}
    \hat{A}(t) = Ue^{\Lambda t}U^{-1}\hat{A}(t=0) - U \left( \frac{e^{\Lambda t} - \mathbb{1}_4}{\Lambda} \right)U^{-1} \sqrt{K} \hat{A}_{\text{in}}.
\end{equation}
We then determine $\langle N_i (t) \rangle = \langle \hat{A}_{i+M}(t) \hat{A}_i(t) \rangle$, the mean photon number of mode $i$, in two different cases.
\begin{itemize}
    \item If $\forall t, \bs{\sigma} (t) = \bs{\sigma}_0$, then
    \begin{equation}
    \label{eq:N_i_t_sigma0}
    \begin{split}
        \langle N_i (t) \rangle &= |\bs{\alpha}_i (t)|^2 \\
        \langle N_i (t)\rangle &= \left|\sqrt{\kappa} \epsilon \sum_{k,m=1}^{2M} U_{im} U^{-1}_{mk} \frac{e^{\lambda_m t} - 1}{\lambda_m} \right|^2
    \end{split}
    \end{equation}
    \item Else the expression is harder to compute
    \begin{equation}
    \label{eq:N_i_t_general}
        \langle N_i (t) \rangle = \kappa |\epsilon|^2 \sum_{k',m',k,m=1}^{2M} U_{i+M,m'}U^{-1}_{m',k'}U_{i,m}U^{-1}_{m,k} \frac {1 + e^{(\lambda_m + \lambda_{m'})t} - e^{\lambda_m t} - e^{\lambda_{m'} t}}{\lambda_{m'}\lambda_m }
    \end{equation}
\end{itemize}
From these formulas, we can infer behaviors of the terms of Eqs.~\eqref{eq:N_i_t_sigma0},\eqref{eq:N_i_t_general} depending on the eigenvalues $\lambda_j$:
\begin{equation}
    \begin{cases}
        \text{Im}(\lambda_j) \neq 0 &\rightarrow \text{ oscillation term}\\
        \text{Re}(\lambda_j) > 0 &\rightarrow \text{ term increasing exponentially in } t\\
        \text{Re}(\lambda_j) < 0 &\rightarrow \text{ term decreasing exponentially in } t\\
    \end{cases}.
\end{equation}

\subsection{2 modes, \texorpdfstring{$g\in \mathbb{R}, g^s=0$}{g in R, gs=0}}
\begin{figure}[h]
    \centering
    \includegraphics[width=0.5\linewidth]{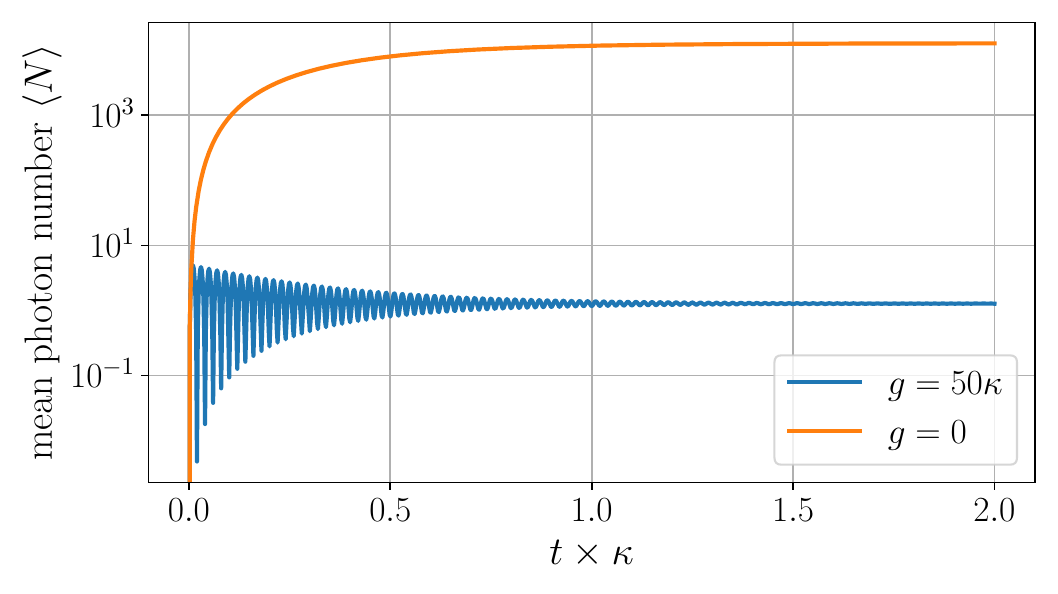}
    \caption{Mean photon number $\langle N (t) \rangle$ as a function of time for 2 modes coupled at photon conversion rates $g=50 \kappa$ and $g=0$. There is no two-mode squeezing, and the drive amplitude $\epsilon$ and dissipation $\kappa$ are defined as in Eq.~\eqref{eq:N_study_default_params}.}
    \label{fig:N_evol_M=2_gs=0_g_influence}
\end{figure}

There is a single coherent coupling process, at a rate $g \in \mathbb{R}$. In this case $\forall t, \bs{\sigma}(t) = \bs{\sigma}_0$, and $F'$ is two-fold degenerate with the diagonalization
\begin{equation}
\begin{split}
    \lambda_\pm &= \pm i g - \frac{\kappa}{2} \\
    \Lambda &= \text{diag}(\lambda_-, \lambda_+, \lambda_-, \lambda_+) \\
    U &= \frac{1}{\sqrt{2}} \begin{pmatrix}
        1 & 1 & 0 & 0\\
        -1 & +1 & 0 & 0\\
        0 & 0 & 1 & -1\\
        0 & 0 & 1 & +1
    \end{pmatrix}\\
\end{split}
\end{equation}
By symmetry, the mean photon numbers in modes $1$ and $2$ are equal: $\langle N_1 \rangle = \langle N_2 \rangle = \langle N \rangle$. We then compute $\langle N \rangle$ using Eq.~\eqref{eq:N_i_t_sigma0}
\begin{equation}
\label{eq:N_t_g_in_R_gs_0}
    \langle N(t) \rangle = \kappa |\epsilon|^2 \times \left( \frac{1 + e^{-\kappa t} - 2 \cos(gt)e^{-\frac{\kappa}{2}t}}{ \left( \frac{\kappa}{2} \right)^2 + g^2} \right).
\end{equation}
The oscillation amplitude of $\langle N \rangle$ being inversely proportional to $(\frac{\kappa}{2})^2 + g^2$ means that decreasing the photon conversion rate $g$ increases the mean photon number $\langle N \rangle$. Figure~\ref{fig:N_evol_M=2_gs=0_g_influence} shows that the average number of photons is $10^4$ larger for $g = \SI{0}{Hz}$ than for $g=50 \kappa$.

\subsection{2 modes, \texorpdfstring{$g \in \mathbb{C}, g^s \in \mathbb{C}$}{g in C, gs in C}}
\begin{figure}[h]
    \centering
    \includegraphics[width=0.5\linewidth]{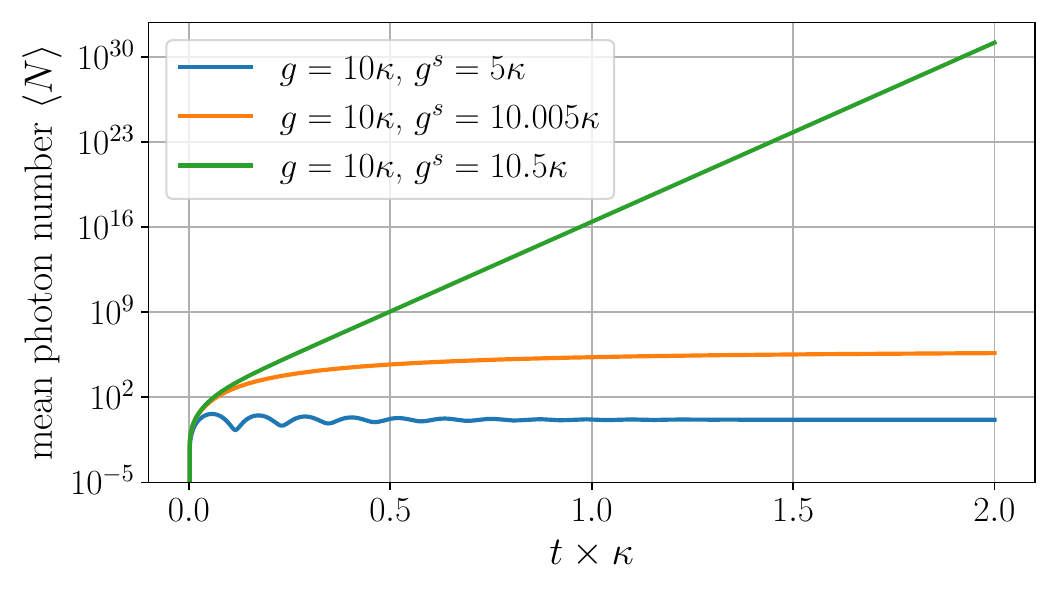}
    \caption{Mean photon number $\langle N (t) \rangle$ for 2 modes as a function of time for $|g^s|<|g|$ (blue) , $|g^s| > |g| \text{ and }2\sqrt{|g^s|^2 - |g|^2} < \kappa$ (orange), $|g^s| > |g| \text{ and }2\sqrt{|g^s|^2 - |g|^2} > \kappa$ (green). The photon conversion rate is fixed at $g=10 \kappa$, and the drive amplitude $\epsilon$ and dissipation $\kappa$ are defined as in Eq.~\eqref{eq:N_study_default_params}.}
    \label{fig:N_evol_M=2_gs_influence}
\end{figure}

We assume both complex-valued photon conversion rate $g$ and two-mode squeezing rate $g^s$. In this case,
\begin{equation}
    F' = \begin{pmatrix}
        0 & -ig & 0 & -ig^s \\
        -ig^* & 0 & -ig^s & 0 \\
        0 & i{g^s}^* & 0 & ig^* \\
        i{g^s}^* & 0 & ig & 0
    \end{pmatrix} - \frac{\kappa}{2}.
\end{equation}
Its diagonalization $F'= U \Lambda U^{-1}$ yields
\begin{align}
    \label{eq:F'_eigencomp}
    \Lambda &= \text{diag}(\lambda_+, \lambda_+, \lambda_-, \lambda_-) \\
    U &= \frac{1}{\mathcal{N}}\begin{pmatrix}
        i(|g^s|^2-|g|^2) & -g & i(|g^s|^2-|g|^2) & -g \\
        g^*\lambda_+ & -i\lambda_+ & g^*\lambda_- & -i\lambda_- \\
        0 & {g^s}^* & 0 & {g^s}^* \\
        -{g^s}^*\lambda_+ & 0 & -{g^s}^*\lambda_- & 0
    \end{pmatrix},
\end{align}
with $\mathcal{N}$ a normalization factor and
\begin{align}
    \lambda_{g, g^s} &= \begin{cases}
    -i\sqrt{|g|^2 - |g^s|^2} &\text{if $|g|>|g^s|$}\\
    \sqrt{|g^s|^2 - |g|^2} &\text{if $|g^s|>|g|$}\\
    \end{cases} \\
    \lambda_\pm &= \pm\lambda_{g, g^s} - \frac{\kappa}{2}.
\end{align}
If $|g| = |g^s|$ i.e the photon conversion rate and two-mode squeezing rate have equal absolute values, then $F'$ is not diagonalizable. By symmetry, the mean photon numbers in mode $1$ and $2$ are equal: $\langle N_1 \rangle = \langle N_2 \rangle = \langle N \rangle$. Then from Eq.~\eqref{eq:N_i_t_general} we get
\begin{equation}
    \langle N(t) \rangle = \kappa |\epsilon|^2 \sum_{k',m',k,m=1}^{2M} U_{1+M,m'}U^{-1}_{m',k'}U_{1,m}U^{-1}_{m,k} \frac {1 + e^{(\lambda_m + \lambda_{m'})t} - e^{\lambda_m t} - e^{\lambda_{m'} t}}{\lambda_{m'}\lambda_m }
\end{equation}
The full expression is too heavy to compute as there are $4^4=256$ terms, but depending on the ratio of the photon conversion and two-mode squeezing rates $g$ and $g^s$, we can infer the behavior of $\langle N (t)\rangle$:
\begin{equation}
\begin{cases}
    \text{if }|g^s| > |g| \text{ and }2\sqrt{|g^s|^2 - |g|^2} > \kappa &\Rightarrow \langle N(t) \rangle \text{ diverges when } t \rightarrow \infty\\
    \text{if }|g^s| > |g| \text{ and }2\sqrt{|g^s|^2 - |g|^2} < \kappa &\Rightarrow \langle N(t) \rangle \text{ doesn't oscillate and converges when } t \rightarrow \infty\\
    \text{if }|g^s| < |g| &\Rightarrow \langle N(t) \rangle \text{ oscillates at frequency } 2\sqrt{|g|^2 - |g^s|^2} \text{ and converges when } t \rightarrow \infty.\\
\end{cases}
\end{equation}
These three different behaviors are shown in Figure~\ref{fig:N_evol_M=2_gs_influence} in the blue, orange and green lines respectively. We observe that if the coherent photon conversion rate and the dissipation are not high enough, the two-mode squeezing tone creates photons at an exponential rate.

\subsection{3 modes, \texorpdfstring{$g \in \mathbb{C}, g^s=0$}{g in C, gs=0}}
Let the photon conversion rates all have equal absolute values, but different phases i.e $g_{kl} = g e^{i\phi^g_{kl}}$ for $k,l \in [1, 2, 3]$, with $g \in \mathbb{R}^+$. The two-mode squeezing tones are turned off. Then
\begin{equation}
\begin{split}
    G & = \begin{pmatrix}
        0 & g_{12} & g_{13}\\
        g_{12}^* & 0 & g_{23}\\
        g_{13}^* & g_{23}^* & 0
    \end{pmatrix} \\
    \mathcal{L} & = \frac{1}{i\hbar} \left(\begin{array}{@{}c|c@{}}
      G
      & \mathbb{0} \\
    \hline
      \mathbb{0} &
      -G^{T}
    \end{array}\right)
\end{split}.
\end{equation}
To compute the eigenvalues of $F' = \mathcal{L} - \frac{\kappa}{2}\mathbb{1}_6$, we compute those of $G$:
\begin{equation}
\begin{split}
    \det(\lambda \mathbb{1} - G ) & = \lambda^3 + p\lambda + q \\
    & \text{with} \begin{cases}
        p & = -3g^2 \\
        q & = -2 g^3 \cos(\phi^g) \\
        \phi^g &= \phi_{12} + \phi_{23} - \phi_{13}
    \end{cases}
\end{split}.
\end{equation}
We solve this cubic equation with Cardano's formula. The discriminant is
\begin{equation}
    \Delta = -(4p^3+27q^2) = 108 g^6 \sin^2(\phi^g).
\end{equation}
We will consider two cases for this discriminant. In the first case, where $\Delta = 0 (\phi^g \equiv 0[\pi])$, there are 3 real eigenvalues of $G$, one simple and a double:
\begin{equation}
\begin{cases}
    \lambda_1 & = 2g \\
    \lambda_2 &= \lambda_3 = -g
\end{cases}.
\end{equation}
The diagonalization of $F'=U\Lambda U^{-1}$ thus yields
\begin{equation}
\begin{cases}
    \Lambda & = \text{diag} (-2ig, ig, ig, 2ig, -ig, -ig) - \frac{\kappa}{2} \mathbb{1}_6 \\
    U & = \frac{1}{\sqrt{6}}\left( \begin{array}{ccc|ccc}
        \sqrt{2} & \sqrt{3} & 1 & 0 & 0 & 0\\
        \sqrt{2} & -\sqrt{3} & 1 & 0 & 0 & 0 \\
        \sqrt{2} & 0 & 2 & 0 & 0 & 0 \\ 
        \hline
        0 & 0 & 0 & \sqrt{2} & \sqrt{3} & 1\\
        0 & 0 & 0 & \sqrt{2} & \sqrt{3} & 1  \\
        0 & 0 & 0 & \sqrt{2} & 0 & 2\\ 
    \end{array} \right)
\end{cases}
\end{equation}
By symmetry, the mean photon number is the same in all modes: $\langle N_1 \rangle = \langle N_2 \rangle = \langle N_3 \rangle = \langle N \rangle$. Using Eq.~\eqref{eq:N_i_t_sigma0} we get the mean photon number $\langle N \rangle$
\begin{equation}
    \langle N(t) \rangle = \kappa |\epsilon|^2 \times \left( \frac{1 + e^{-\kappa t} - 2 \cos(2gt)e^{-\frac{\kappa}{2}t}}{ \left( \frac{\kappa}{2} \right)^2 + (2g)^2} \right)
\end{equation}
This result is the same as Eq.~\eqref{eq:N_t_g_in_R_gs_0}, but by substituting $g \rightarrow 2g$.

\begin{figure}[h]
    \centering
    \includegraphics[width=0.5\linewidth]{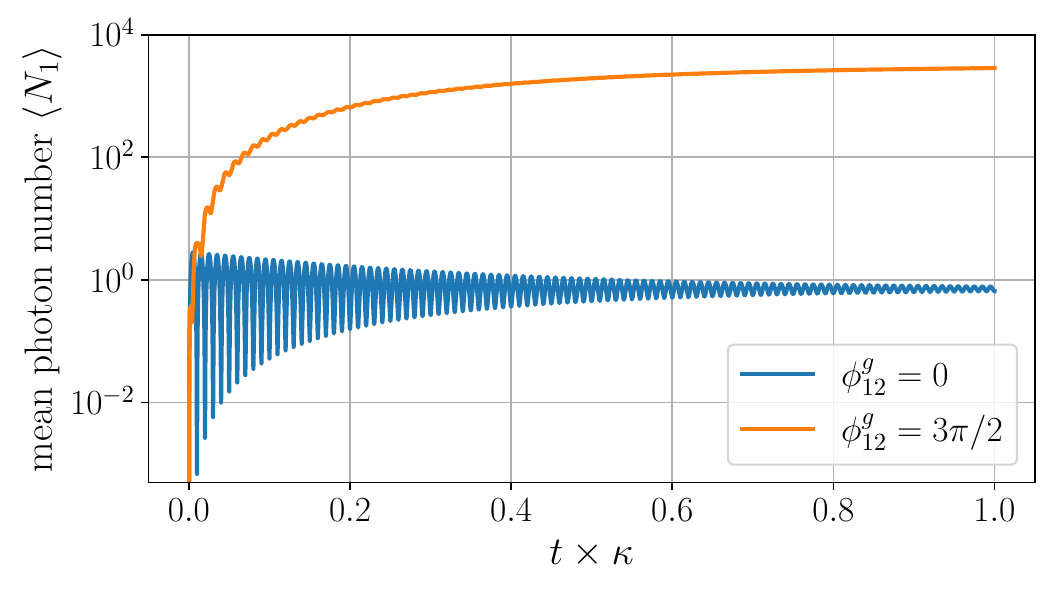}
    \caption{Mean photon number $\langle N_1 (t) \rangle$ as a function of time for 3 modes, for photon conversion rate phases $\phi^g_{12}=0$ (blue) and $\phi^g_{12}=\frac{3\pi}{2}$ (orange), while the others are $\phi^g_{13} = \phi^g_{23} = 0$. The conversion rates amplitude are $|g_{kl}|=g=50 \kappa$ and there is no two-mode squeezing.}
    \label{fig:N_evol_M=3_g_phase_influence}
\end{figure}

In the second case where $\Delta > 0 (\phi^g \not\equiv 0[\pi])$, there are 3 real degenerate eigenvalues of $G$
\begin{equation}
\label{eq:eigvals_3_modes_cos}
\begin{split}
    \lambda_{k+1} = 2g \cos\left( \frac{|\phi^g|}{3} + \frac{2k\pi}{3} \right) & \qquad \begin{cases}
        k \in [0, 1, 2] \\
        \phi^g \in \{ -\pi,\pi \}
    \end{cases}
\end{split}.
\end{equation}
The full formula of the mean photon number in mode $i$ $\langle N_i \rangle$ is too big to compute, but we observe that if $|\phi^g| \equiv \frac{(3 - 4k)\pi}{2} [3\pi]$, then the eigenvalue $\lambda_k$ becomes null. Hence the term associated with this eigenvalue will not oscillate, and its value evolves as if there had been no photon conversion i.e it is divided by $\kappa$ and not by a term $\approx  g^2 + \left( \frac{\kappa}{2}\right)^2$, leading to much higher $\langle N_i (t) \rangle$. This behavior is illustrated in Figure~\ref{fig:N_evol_M=3_g_phase_influence} for $|\phi^g| = \frac{3\pi}{2}$. We can interpret this as a destructive inference between different photon conversion processes, leading to the average photon number $\langle N_i (t) \rangle$ that has a similar dynamics to the $g=0$ case.

\subsection{3 modes, \texorpdfstring{$g\neq0, g^s\neq0$}{g!=0, gs!=0}}
\begin{figure}[h]
    \centering
    \vspace{5pt}%
    \begin{subfigure}{\textwidth}
        \centering
        \begin{minipage}[t]{0.05\textwidth}
            \vspace{-60pt}
            \caption{\label{fig:N_cmap_a}}
        \end{minipage}\hfill
        \begin{minipage}[c]{0.90\textwidth}
            \includegraphics[width=\textwidth]{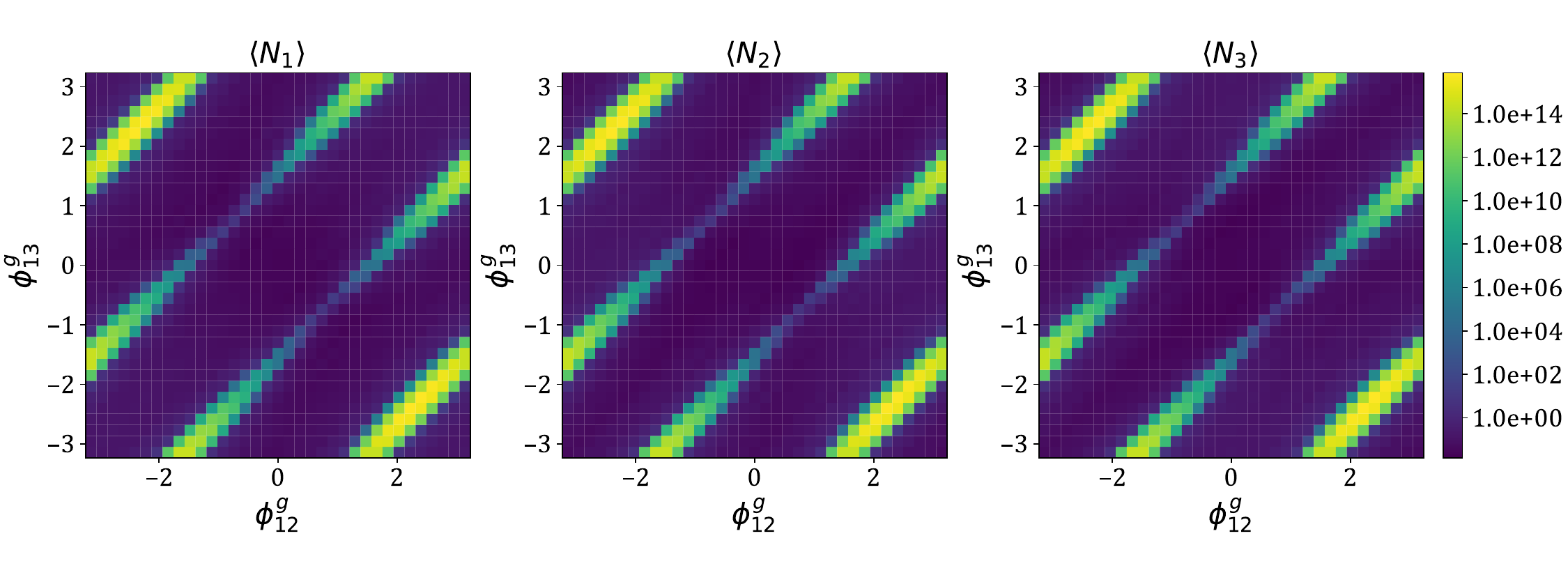}
        \end{minipage}
    \end{subfigure}
    \par\vspace{5pt}
    \begin{subfigure}{\textwidth}
        \centering
        \begin{minipage}[t]{0.05\textwidth}
            \vspace{-60pt}
            \caption{\label{fig:N_cmap_b}}
        \end{minipage}\hfill
        \begin{minipage}[c]{0.90\textwidth}
            \includegraphics[width=\textwidth]{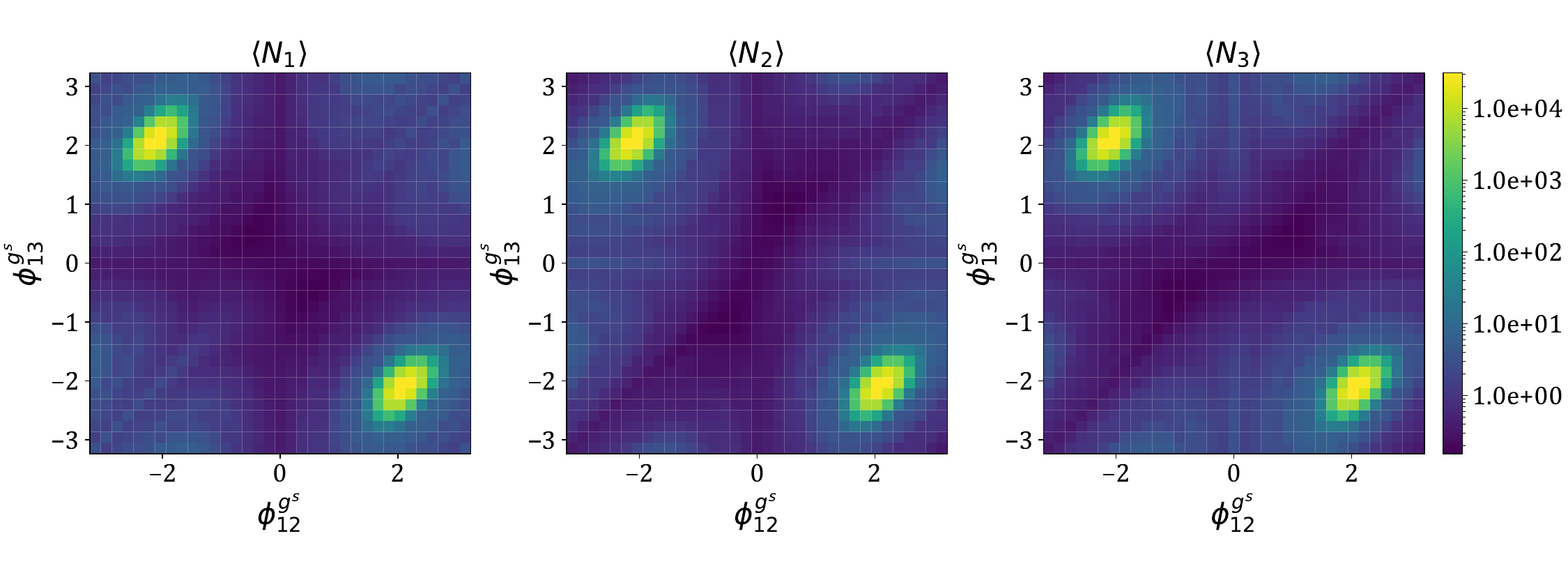}
        \end{minipage}
    \end{subfigure}
    \par\vspace{5pt}
    \begin{subfigure}{\textwidth}
        \centering
        \begin{minipage}[t]{0.05\textwidth}
            \vspace{-60pt}
            \caption{\label{fig:N_cmap_c}}
        \end{minipage}\hfill
        \begin{minipage}[c]{0.90\textwidth}
            \includegraphics[width=\textwidth]{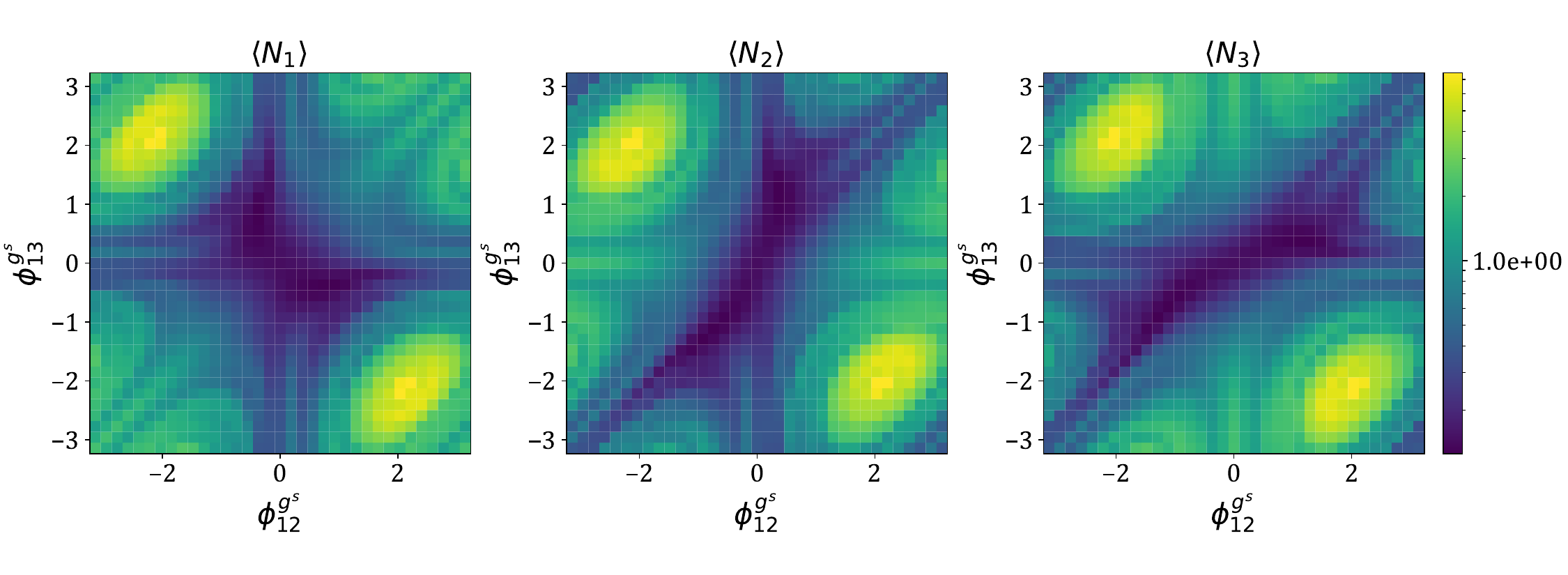}
        \end{minipage}
    \end{subfigure}

    \caption{(a) The mean photon number $\langle N_i (t) \rangle$ in 3 modes as a function of photon conversion rate phases $\phi^{g}_{01}$ and $\phi^{g}_{02}$ at time $t=0.4 \pi \kappa^{-1}$, the photon conversion rates all have the same absolute value $|g_{kl}|=g=50 \kappa$ and the two-mode squeezing rates are ${g^s}_{kl}=g^s=10 \kappa$. (b)(c) The mean photon number $\langle N_i (t) \rangle$ in the mode $i$ as a function of the two-mode squeezing rate phases $\phi^{g^s}_{01}$ and $\phi^{g^s}_{02}$ at time $t=0.4 \pi \kappa^{-1}$ for  $|g_{kl}|=g=19.5 \kappa$ and $|{g^s}_{kl}| = g^s=10 \kappa$ (b), or $g=20.5 \kappa$ and $g^s=10 \kappa$ (c).}
    \label{fig:N_cmap}
\end{figure}

Computing the eigenvalues of $F'$ to compute the mean photon number in Eq.~\eqref{eq:N_i_t_general} in this case is not possible, so we resort to numerical simulations. There are different dynamical regimes depending on the photon conversion and two-mode squeezing rates. We illustrate these regimes in Figure~\ref{fig:N_cmap}. In all of the studies listed below, all the photon conversion rates are identical $|g_{kl}| = g$, as well as the two-mode squeezing rates $|{g^s}_{kl}| = g^s$. We show that, depending on the ratio of different coupling rates, certain combinations of coupling rate phases lead to photon number divergence.
\begin{itemize}
    \item Figure~\ref{fig:N_cmap_a} shows the mean photon number as a function of the photon conversion rate phases $\phi^{g}_{01}$ and $\phi^{g}_{02}$. We observe that for certain phase values the number of photons diverges. We interpret this as a consequence of the destructive interference between multiple photon conversion processes, resulting in the creation of photons through two-mode squeezing processes, and the lack of effective photon conversion to dampen them.
    
    \item Figure~\ref{fig:N_cmap_b} shows the mean photon number as a function of two two-mode squeezing rate phases $\phi^{g^s}_{01}$ and $\phi^{g^s}_{02}$, for $g < 2 g^s$ . We observe that for certain phase values the number of photons diverges. We interpret this as a consequence of constructive interference of the two-mode squeezing processes, resulting in an effective two-mode squeezing rate $2 g^s$ that is higher than the photon conversion rate, so that the number of photons diverges.
    
    \item Figure~\ref{fig:N_cmap_c} shows the mean photon number as a function of two two-mode squeezing rate phases $\phi^{g^s}_{01}$ and $\phi^{g^s}_{02}$, for $g > 2 g^s$ . We observe that the number of photons never diverges, although it is higher for phase values that cause the case $g < 2 g^s$ to diverge. We interpret this as constructive interferences in two-mode squeezing tones not having high enough coupling rates to surpass photon conversion rates, which dampens photon creation.
\end{itemize}

\section{Clamping of the coupling parameters}
\label{sec:clamping}
Taking into account the behavior of the photon numbers $\langle N_i (t) \rangle$ with respect to the photon conversion and two-mode squeezing rates observed in Section~\ref{sec:divN} , we propose a set of heuristic guidelines for choosing the coupling rates to avoid exponential divergences in the photon numbers of the bosonic modes. We stress that these guidelines are not rigorously proven methods for preventing such divergences, but rather practical intuitions that have been effective in training the coupling parameters.
\begin{algorithm}[H] 
\caption{Clamping general guidelines}
\begin{algorithmic}[1]
    \State Clamping is applied element-wise to each component of the coupling rates.
    \State Photon conversion rates are restricted to real, positive values. Complex values may result in destructive interference during photon conversion. In contrast, two-mode squeezing rates can be complex-valued.
    \State The amplitude of the highest two-mode squeezing rate should never be higher than the amplitude of the lowest photon conversion rate.
    \State If the input is encoded in the phase of the two-mode squeezing rates of $M$ modes, then the highest two-mode squeezing amplitude should never be higher than the lowest photon conversion rate amplitude divided by $M-1$.
    \If{the input $\bf{x}$ is encoded in one of the coupling rates according to Eq. (2)}
        \State If the encoding variable $\bs{\theta}(\bf{x})$ requires clamping after a gradient descent update, the bias term $\bs{\theta}_{\text{bias}}$  is adjusted to enforce the clamping constraints, while $\bs{\theta}_0$ remains fixed. If this is insufficient to satisfy the clamping conditions, $\bs{\theta}_0$ is also clamped.
        \State  The values $\bs{\varphi}_0$ and $\bs{\varphi}_{\text{bias}}$ are never clamped, even when phase encoding is used.
        \State After clamping, the explored values of $\bs{\theta}(\bf{x})$ should deviate as little as possible from their original (pre-clamping) values.
    \EndIf
\end{algorithmic}
\end{algorithm}
Following these guidelines, we propose an algorithm to clamp the photon conversion rates $\bs{g}$ and the two-mode squeezing rates $\bs{g^s}$. We define the clamping bounds $l^g_{\min}, l^g_{\max} \in \mathbb{R}^+$ for $\bs{g}$, and  $l^{g^s}_{\min}, l^{g^s}_{\max} \in \mathbb{R}^+$ for $\bs{g^s}$.
\begin{algorithm}[H] 
\caption{Clamping rules}\label{alg:clamping_rules}
\begin{algorithmic}
    \Require $\bs{g} \in \left( \mathbb{R}^+ \right)^{\frac{M(M-1)}{2}}$ \Comment{To prevent destructive interference in the photon conversion rates}
    \Require $\bs{g^s} \in  \mathbb{C}^{\frac{M(M-1)}{2}}$ \Comment{The squeezing rates are allowed to be complex.}
    \State Clamping is applied element-wise to each component of the coupling rates.
    \State $l^{g^s}_{\min} \gets \SI{0}{Hz}$
    \State $l^{g}_{\max} \gets \SI{500}{MHz}$
    \If{the input $\bf{x}$ is encoded in $\arg(\bs{g^s})$}
    \Comment{We require that $\max(|\bs{g^s}|) < \frac{\min(\bs{g})}{M-1}$}
        \State $l^{g^s}_{\max} \gets$ arbitrary constant value
        \State $l^g_{\min} \gets l^{g^s}_{\max} \times (M-1)$
    \Else
    \Comment{We require that $\max(|\bs{g^s}|) < \min(\bs{g})$}
        \State $l^g_{\min} \gets \frac{\max(\bs{g^s}) + \min(\bs{g})}{2}$
        \State $l^{g^s}_{\max} \gets \frac{\max(\bs{g^s}) + \min(\bs{g})}{2}$
    \EndIf
    \If{the input $\bf{x}$ is encoded in $\bs{\epsilon}$}
        \State $\bs{g} \gets \text{clamp}(\bs{g}, l^g_{\min}, l^g_{\max})$
        \State $\bs{|g^s|} \gets \text{clamp}(\bs{|g^s|}, 0, l^{g^s}_{\max})$
    \ElsIf{the input $\bf{x} \in [0, 1]$ is encoded in $\bs{g^s}$ using the equation $\bs{g^s}(\bf{x}) = \bs{\mathnormal{g}^s}_0 \cdot \bf{x} + \bs{\mathnormal{g}^s}_{\text{bias}}$}
        \Comment{$\bs{g}_0, \bs{g}_{\text{bias}} \in (\mathbb{R}^+)^{\frac{M(M-1)}{2}}$}
        \State $|\bs{g^s}| \gets \text{clamp}(|\bs{g^s}|, 0, l^{g^s}_{\max})$
        \State $\bs{g}_0 \gets \text{clamp}(\bs{g}_0, 0, l^{g}_{\max} - l^{g}_{\min})$
        \State $\bs{g}_{\text{bias}} \gets \text{clamp}(\bs{g}_{\text{bias}}, 0, l^{g}_{\min})$
    \ElsIf{the input $\bf{x} \in [0, 1]$ is encoded in $\bs{g^s}$ using the equation $\bs{g^s}(\bf{x}) = \bs{\mathnormal{g}^s}_0 \cdot e^{i (\varphi_0 \cdot \bf{x} + \varphi_{\text{bias}})} + \bs{\mathnormal{g}^s}_{\text{bias}}$}
        \State $\bs{g} \gets \text{clamp}(\bs{g}, l^g_{\min}, l^g_{\max})$
        \State $|\bs{g^s}_0| \gets \text{clamp}(|\bs{g^s}_0|, 0, l^{g^s}_{\max})$
        \State $|\bs{g^s}_{\text{bias}}| \gets \text{clamp}(|\bs{g^s}_{\text{bias}}|, 0, l^{g^s}_{\max} - |\bs{g^s}_0|)$
    \ElsIf{the input $\bf{x} \in [0, 1]$ is encoded in $\bs{g^s}$ using the equation $\bs{g^s}(\bf{x}) = \bs{\mathnormal{g}^s}_0 \cdot \bf{x} + \bs{\mathnormal{g}^s}_{\text{bias}}$}
        \State $\bs{g} \gets \text{clamp}(\bs{g}, l^g_{\min}, l^g_{\max})$
        \State $|\bs{g^s}_0| \gets \text{clamp}(|\bs{g^s}_0|, 0, l^{g^s}_{\max})$
        \If{there exists any $\mathbf{x} \in [0, 1]$ such that $|\bs{g^s}(\mathbf{x})| \notin [0, l^{g^s}_{\max}]$}
            \State $\bs{g^s}_{\text{bias}}$ is modified such that $|\bs{g^s}(\mathbf{x})| \in [0, l^{g^s}_{\max}]$ for all $\mathbf{x} \in [0, 1]$
            \LComment{The updated values of $\bs{g^s}(\mathbf{x})$ should deviate as little as possible from their original (pre-clamping) values. The detailed clamping procedure is implemented in the source code, specifically in:\newline\textbf{tests/clamping\_demonstrations/abs\_encoded\_cplx\_theta\_clamp.ipynb}}
        \EndIf
    \EndIf
\end{algorithmic}
\end{algorithm}

Where $\bs{p} \rightarrow \text{clamp}(\bs{p}, p_{\min}, p_{\max})$ denotes the element-wise operation defined as $p_i \rightarrow \min(\max(p_i, p_{\min}), p_{\max})$ for each element $p_i$ of the vector $\bs{p} \in \mathbb{R}^{s_{\bs{p}}}$. In practice, we find that these clamping rules effectively prevent divergences in the number of photons across all learning tasks.

\subsection{Number of trainable physical parameters}
As a result of the clamping rules, the exchange coupling rates are constrained to real values. So counting the number of trainable physical parameter yields:
\begin{equation}
\begin{cases}
    \bs{\epsilon} \in \mathbb{C}^M &\rightarrow 2M \\
    \bs{\delta} \in \mathbb{R}^M &\rightarrow M \\
    \bs{g} \in \mathbb{R}^{\frac{M(M-1)}{2}} &\rightarrow \frac{M(M-1)}{2} \\
    \bs{g^s} \in \mathbb{C}^{\frac{M(M-1)}{2}} &\rightarrow M(M-1), \\
\end{cases}
\end{equation}
adding up to $\frac{3}{2}M(M-1) + 3M$.
\section{Gradient of the Loop Hafnian}

Gradient of Fock state occupation probabilities can be computed analytically by computing the gradient of loop Hafnians. Even though in practice we use numerical gradients obtained with the backpropagation algorithm,
in this section we show the analytical calculation. 

We consider a system of $M$ modes, whose Gaussian state is characterized by a displacement vector $\bs{\alpha}$ and a covariance matrix $\bs{\sigma}$, of dimensions $2M$ and $2M \times 2M$, respectively. Both $\bs{\alpha}$ and $\bs{\sigma}$ depend on a parameter $\theta$. Our goal is to compute the derivative $\partial_{\theta} \mathrm{lhaf}(\bs{A}_{\bar{n}})$, as defined in the Methods section, following the approach outlined in~\cite{banchiTrainingGaussianBoson2020}. According to Wick’s theorem,
\begin{equation}
    \mathrm{lhaf}(\bs{A}_{\bar{n}}, \bs{\gamma}_{\bar{n}})=\int \prod_{j=1}^M dx_j \frac{e^{-\frac{1}{2}(x-\bs{\gamma})^T \bs{A}^{-1} (x-\bs{\gamma})}}{\sqrt{\mathrm{det}(2\pi \bs{A}})} x_1^{n_1} \dots x_M^{n_M}.
\end{equation}
We differentiate the exponential term in the integral with respect to $\theta$
\begin{equation}
    \partial_\theta \left( (x-\bs{\gamma})^T \bs{A}^{-1} (x-\bs{\gamma}) \right) = -2(\partial_\theta \bs{\gamma})^T \bs{A}^{-1} (x-\bs{\gamma}) - (x-\bs{\gamma})^T \bs{A}^{-1} (\partial_\theta \bs{A}) \bs{A}^{-1} (x-\bs{\gamma}).
\end{equation}
So the exponential term becomes
\begin{equation}
\begin{split}
    \partial_\theta (\mathrm{lhaf}(\bs{A}_{\bar{n}}, \bs{\gamma}_{\bar{n}})) & = \frac{1}{2} \sum_{k,l} (\bs{A}^{-1} (\partial_\theta \bs{A}) \bs{A}^{-1})_{k,l} \int \prod_{j=1}^M dx_j \frac{e^{-\frac{1}{2}(x-\bs{\gamma})^T \bs{A}^{-1} (x-\bs{\gamma})}}{\sqrt{\mathrm{det}(2\pi \bs{A}})} (x-\bs{\gamma})_k(x-\bs{\gamma})_l x_1^{n_1} \dots x_M^{n_M}\\
    & + \sum_{k,l} (\partial_\theta \bs{\gamma})_k (\bs{A}^{-1})_{k,l} \int \prod_{j=1}^M dx_j \frac{e^{-\frac{1}{2}(x-\bs{\gamma})^T \bs{A}^{-1} (x-\bs{\gamma})}}{\sqrt{\mathrm{det}(2\pi \bs{A}})} (x-\bs{\gamma})_l x_1^{n_1} \dots x_M^{n_M}\\
    & - \frac{1}{2} \mathrm{Tr}[\bs{A}^{-1} \partial_\theta \bs{A}] \mathrm{lhaf}(\bs{A}_{\bar{n}}, \bs{\gamma}_{\bar{n}}).
\end{split}
\end{equation}
The integrals are simplified into Hafnian expressions, with the $\bs{\gamma}_{\bar{n}}$ terms in the loop Hafnian omitted for clarity. We also adopt the notation $(\bar{e}_k)_i = \delta_{ik}$.
\begin{equation}
\label{eq:derivative1}
\begin{split}
    \partial_\theta (\mathrm{lhaf}(\bs{A}_{\bar{n}}, \bs{\gamma}_{\bar{n}})) & = \frac{1}{2} \sum_{k,l} (\bs{A}^{-1} (\partial_\theta \bs{A}) \bs{A}^{-1})_{k,l} \left[ \mathrm{lhaf}(\bs{A}_{\bar{n}+\bar{e}_k+\bar{e}_l}) - \bs{\gamma}_l \mathrm{lhaf}(\bs{A}_{\bar{n}+\bar{e}_k}) - \bs{\gamma}_k \mathrm{lhaf}(\bs{A}_{\bar{n}+\bar{e}_l}) + \bs{\gamma}_l\bs{\gamma}_k\mathrm{lhaf}(\bs{A}_{\bar{n}})\right]\\
    & + \sum_{k,l} (\partial_\theta \bs{\gamma})_k (\bs{A}^{-1})_{k,l} \left[ \mathrm{lhaf}(\bs{A}_{\bar{n}+\bar{e}_l}) - \bs{\gamma}_l \mathrm{lhaf}(\bs{A}_{\bar{n}}) \right]\\
    & - \frac{1}{2} \mathrm{Tr}[\bs{A}^{-1} \partial_\theta \bs{A}] \mathrm{lhaf}(\bs{A}_{\bar{n}}).
\end{split}
\end{equation}
The Laplace-like expansion of the Hafnian (for fixed $c$) is
\begin{equation}
    \mathrm{Haf}(\bs{A}) = \sum_{j\neq c} \bs{A}_{jc} \mathrm{Haf}(\bs{A}_{-j-c}), 
\end{equation}
where $\bs{A}_{-j-c}$ denotes the matrix $\bs{A}$ with rows and columns $j$ and $c$ removed. This can be understood by considering the enumeration of Perfect Pair Matchings when a vertex is added to a graph.  
A similar expansion can be derived for the loop Hafnian by including the single-loop term. For a fixed index $c$, we obtain
\begin{equation}
\label{eq:Laplace_expansion}
    \mathrm{lhaf}(\bs{A}) = \bs{A}_{cc} \mathrm{lhaf}(\bs{A}_{-c}) + \sum_{j\neq c} \bs{A}_{jc} \mathrm{lhaf}(\bs{A}_{-j-c}).
\end{equation}
Now using Eq.~\eqref{eq:Laplace_expansion}, we get
\begin{equation}
\label{eq:expansion1}
      \mathrm{lhaf}(\bs{A}_{\bar{n}+\bar{e}_k+\bar{e}_l}) = - \bs{A}_{kl} \mathrm{lhaf}(\bs{A}_{\bar{n}}) + \bs{A}_{ll} \mathrm{lhaf}(\bs{A}_{\bar{n}+\bar{e}_k}) + \sum_{j=1}^m \bs{A}_{jl} \mathrm{lhaf}(\bs{A}_{\bar{n}+\bar{e}_k -\bar{e}_j}) 
\end{equation}
\begin{equation}
      \mathrm{lhaf}(\bs{A}_{\bar{n}+\bar{e}_k}) = \bs{A}_{kk} \mathrm{lhaf}(\bs{A}_{\bar{n}}) + \sum_{j\neq k}^m \bs{A}_{jk} \mathrm{lhaf}(\bs{A}_{\bar{n}-\bar{e}_j}) 
\end{equation}
\begin{equation}
      \mathrm{lhaf}(\bs{A}_{\bar{n}+\bar{e}_k-\bar{e}_j}) = \bs{A}_{kk} \mathrm{lhaf}(\bs{A}_{\bar{n}-\bar{e}_j}) + \sum_{i \neq k}^m \bs{A}_{ik} \mathrm{lhaf}(\bs{A}_{\bar{n}-\bar{e}_j-\bar{e}_i}).
\end{equation}

We can eliminate the trace term in Eq.~\eqref{eq:derivative1}, using the $\bs{A}_{kl}$ term in Eq.~\eqref{eq:expansion1}
\begin{equation}
    \begin{split}
        \frac{1}{2} \sum_{kl} (\bs{A}^{-1}(\partial_\theta \bs{A})\bs{A}^{-1})_{kl} \bs{A}_{kl} \mathrm{lhaf}(\bs{A}_{\bar{n}}) &= \frac{1}{2} \sum_{kl} \sum_{rs} \bs{A}_{kr}^{-1} (\partial_\theta \bs{A})_{rs} \bs{A}_{sl}^{-1} \bs{A}_{kl} \mathrm{lhaf}(\bs{A}_{\bar{n}}) \\
        &= \frac{1}{2} \sum_{k} \sum_{r} \bs{A}_{kr}^{-1} (\partial_\theta \bs{A})_{rk} \mathrm{lhaf}(\bs{A}_{\bar{n}}) \\
        &= \frac{1}{2} \mathrm{Tr}[\bs{A}^{-1} \partial_\theta \bs{A}] \mathrm{lhaf}(\bs{A}_{\bar{n}}).
    \end{split}
\end{equation}
We can prove that the gradient of the loop hafnian is then
\begin{equation}
    \label{eq:gradient_loophafnian}
    \partial_\theta \mathrm{lhaf}(\bs{A}_{\bar{n}}, \bs{\gamma}_{\bar{n}}) = \frac{1}{2} \sum_j \sum_{i\neq j} (\partial_\theta \bs{A}_{\bar{n}})_{ij} \mathrm{lhaf}(\bs{A}_{\bar{n}-\bar{e}_j-\bar{e}_i}) + \sum_j (\partial_\theta \bs{\gamma})_j \mathrm{lhaf}(\bs{A}_{\bar{n}-\bar{e}_j}).
\end{equation}
Knowing $\bs{\sigma}_Q = (\mathbb{1}_{2M} - \bs{T}\bs{A})^{-1}$, we get from ref.~\cite{banchiTrainingGaussianBoson2020}
\begin{equation}
    \partial_\theta \left( \frac{1}{\sqrt{\mathrm{det}(\bs{\sigma}_Q)}} \right) = -\frac{1}{2} \mathrm{Tr} \left[ \sqrt{\mathrm{det}(\mathbb{1}_{2M}-\bs{T}\bs{A})} \frac{\partial_\theta \bs{A}}{\bs{T}-\bs{A}}  \right].
\end{equation}
We differentiate the exponential term of the GBS formula
\begin{equation}
    \partial_\theta(\exp(-\frac{1}{2} \bs{\alpha}^\dagger \bs{\sigma}_Q^{-1} \bs{\alpha})) = \left( -\partial_\theta \bs{\alpha}^\dagger (\mathbb{1}_{2m}-\bs{T}\bs{A})\bs{\alpha} + \frac{1}{2}\bs{\alpha}^\dagger \bs{T} (\partial_\theta \bs{A})\bs{\alpha} \right) \exp(-\frac{1}{2} \bs{\alpha}^\dagger \bs{\sigma}_Q^{-1} \bs{\alpha}).
\end{equation}
We can now compute the final derivative of the GBS formula
\begin{equation}
\label{eq:GBS_derivative}
\begin{aligned}
    \partial_\theta P_{\bs{A}}(\bar{n}) &= -\frac{1}{2} \mathrm{Tr}\left[\frac{\partial_\theta \bs{A}}{\bs{T} - \bs{A}}\right] P_{\bs{A}}(\bar{n}) \\
    &+ P_{\bs{A}}(\bar{n}) \left( -\partial_\theta \bs{\alpha}^\dagger (\mathbb{1}-\bs{T}\bs{A})\bs{\alpha} + \frac{1}{2}\bs{\alpha}^\dagger \bs{T} (\partial_\theta \bs{A})\bs{\alpha} \right) \\
    &+ \frac{1}{\sqrt{\mathrm{det}(\bs{\sigma}_Q)} \prod_i n_i!} \sum_{i\neq j}^{2N} (\partial_\theta \bs{A}_{\bar{n}\oplus\bar{n}})_{ij} \mathrm{lhaf}(\bs{A}_{\bar{n}\oplus\bar{n}}^{[i,j]}) \\
    &+ \frac{1}{\sqrt{\mathrm{det}(\bs{\sigma}_Q)} \prod_i n_i!} \sum_{j=1}^{2N} (\partial_\theta \bs{\gamma})_j \mathrm{lhaf}(\bs{A}_{\bar{n}\oplus\bar{n}}^{[i]}) 
\end{aligned}
.
\end{equation}
with $N = \sum_i n_i$, and the submatrix $\bs{A}_{\bar{n}\oplus\bar{n}}^{[i,j]}$ is obtained from $\bs{A}_{\bar{n}\oplus\bar{n}}$ by removing rows and columns $i$ and $j$.

\section{Calculation of \texorpdfstring{$\bs{\sigma}$}{sigma}(t)} 
\label{sec:sigma_t_calc}

We compute the different terms of the covariance matrix
\begin{equation}
\begin{split}
    \bs{\alpha}_i(t)\bs{\alpha}_j^*(t) &= \sum_{k,l} F_{ik}(t)F_{jl}^*(t) \bs{\alpha}_k(t=0) \bs{\alpha}_l^*(t=0) \\
    & + \sum_{k,l} \int_0^t \int_0^t F_{ik}(t-\tau) F_{jl}^*(t-\tau') \sqrt{K_k K_l}\bs{\alpha}_{\text{in},k}(\tau) \bs{\alpha}^*_{\text{in},l}(\tau') d\tau d\tau'\\
    & - \sum_{k,l} F_{ik}(t) \bs{\alpha}_k(t=0) \int_0^t F_{jl}^*(t-\tau) \sqrt{K_l} \bs{\alpha}_{\text{in},l}^*(\tau) d\tau \\
    & - \sum_{k,l} F_{jl}^*(t) \bs{\alpha}_l^*(t=0) \int_0^t F_{ik}(t-\tau) \sqrt{K_k} \bs{\alpha}_{\text{in},k}(\tau) d\tau, 
\end{split}
\end{equation}
    
\begin{equation}
\begin{split}
    \langle \hat{A}_i \hat{A}_j^\dagger \rangle (t) &= \sum_{k,l} F_{ik}(t) F_{jl}^*(t) \langle \hat{A}_k \hat{A}_l^\dagger \rangle (t=0) \\
    & + \sum_{k,l} \int_0^t \int_0^t F_{ik}(t-\tau) F_{jl}^*(t-\tau') \sqrt{K_k K_l} \langle \hat{A}_{\text{in}, k}(\tau) \hat{A}_{\text{in}, l}^\dagger(\tau') \rangle d\tau d\tau'\\
    & - \sum_{k,l} F_{ik}(t) \bs{\alpha}_k(t=0) \int_0^t F_{jl}^*(t-\tau) \sqrt{K_l} \bs{\alpha}_{\text{in},l}^*(\tau) d\tau \\
    & - \sum_{k,l} F_{jl}^*(t) \bs{\alpha}_l^*(t=0) \int_0^t F_{ik}(t-\tau) \sqrt{K_k} \bs{\alpha}_{\text{in},k}(\tau) d\tau.
\end{split}
\end{equation}
We observe that the two last terms in $\bs{\alpha}_i(t)\bs{\alpha}_j^*(t)$ and $\langle \hat{A}_i \hat{A}_j^\dagger \rangle (t)$ will cancel out. So the expression for the covariance matrix is
\begin{equation}
\begin{split}
    \bs{\sigma}_{ij}(t) &= \sum_{k,l} F_{ik}(t) F_{jl}^*(t) \left( \frac{1}{2} \langle \hat{A}_k \hat{A}_l^\dagger + \hat{A}_l^\dagger \hat{A}_k \rangle (t=0) - \bs{\alpha}_k(t=0) \bs{\alpha}_l^*(t=0) \right) \\
    & + \sum_{k,l} \sqrt{K_k K_l} \int_0^t \int_0^t F_{ik}(t-\tau) F_{jl}^*(t-\tau')  \left( \frac{1}{2} \langle \hat{A}_{\text{in}, k}(\tau) \hat{A}_{\text{in}, l}^\dagger(\tau') + \hat{A}_{\text{in}, l}^\dagger(\tau') \hat{A}_{\text{in}, k}(\tau) \rangle - \bs{\alpha}_{\text{in},k}(\tau) \bs{\alpha}^*_{\text{in},l}(\tau') \vphantom{\frac{1}{2}}\right) d\tau d\tau' \\
    \bs{\sigma}_{ij}(t) & = \sum_{k,l} F_{ik}(t) F_{jl}^*(t) \bs{\sigma}_{kl}(t=0) + \sum_{k,l} \sqrt{K_k K_l} \int_0^t \int_0^t F_{ik}(t-\tau) F_{jl}^*(t-\tau') \bs{\sigma}_{\text{in},kl}(\tau, \tau') d\tau d\tau'.
\end{split}
\end{equation}
Then the final expression for $\bs{\sigma}(t)$ is obtained. 
The input modes $\hat{A}_{\text{in}}$ have coherent states, so $\bs{\sigma}_{\text{in}}(\tau, \tau') = \bs{\sigma}_0 \delta(\tau - \tau')$. We then get
\begin{equation}
    \bs{\sigma}_{ij}(t) = \sum_{k,l} F_{ik}(t) F_{jl}^*(t) \bs{\sigma}_{kl}(t=0) + \sum_{k,l}  \sqrt{K_k K_l} \int_0^t F_{ik}(t-\tau) F_{jl}^*(t-\tau') (\bs{\sigma}_{0})_{kl} d\tau.
\end{equation}
Since $\bs{\sigma}_0 = \frac{1}{2} \mathbb{1}_{2 M}$,
\begin{equation}
    \bs{\sigma}_{ij}(t) = \sum_{k,l} F_{ik}(t) F_{jl}^*(t) \bs{\sigma}_{kl}(t=0) + \frac{1}{2} \sum_{k}  K_k \int_0^t F_{ik}(t-\tau) F_{jk}^*(t-\tau)  d\tau
\end{equation}
\begin{equation}
    \bs{\sigma}(t) = F(t) \bs{\sigma}(t=0) F^\dagger(t) + \frac{1}{2} \int_0^t F(t-\tau) K F^\dagger(t - \tau) d\tau.
\end{equation}

\def\bibsection{\section*{Supplementary References}}
\bibliography{SuppMat.bib}